\documentclass[a4paper,10pt]{article}
\usepackage[utf8]{inputenc}
\usepackage{authblk} 
\usepackage{amsmath, amsthm, amsfonts, amssymb} 
\usepackage{mathrsfs} 
\usepackage{bbm}
\usepackage{mathtools}
\usepackage{doi}
\usepackage{hyperref} 
\usepackage{braket}
\usepackage{IKKtex} 
\numberwithin{equation}{section}

\def\thintablerule{\hrule height0.4pt}
\allowdisplaybreaks

\usepackage[backend=bibtex,sorting=none,citestyle=numeric-comp]{biblatex}
\renewbibmacro{in:}{}
\DeclareFieldFormat{pages}{#1}
\usepackage{subfiles}
\usepackage{blindtext} 
\usepackage{xr-hyper}
\usepackage{refcount}
\usepackage{subfiles}

\begin{document}

\centerline {\Huge The thermal backreaction of a scalar field}
\vskip .5cm
\centerline {\Huge in de Sitter spacetime}

\vskip 2 cm
\centerline{\large Nikos Irges$^1$, Antonis Kalogirou$^1$ and Fotis Koutroulis$^{2,3}$}
\vskip.5cm
\begin{center}
{\it 1. Department of Physics, National Technical University of Athens}\\
{\it Zografou Campus, GR-15780 Athens, Greece}
\end{center}
\vskip.5cm
\begin{center}
{\it 2. Theoretical Physics Division, Institute of High Energy Physics,\\ Chinese Academy of Sciences,}
{\it 19B Yuquan Road,\\ Shijingshan District, Beijing 100049, China}\\
{\it 3. China Center of Advanced Science and Technology, Beijing 100190, China}
\end{center}
\vskip .4cm
\begin{center}
{\it e-mail: irges@mail.ntua.gr, akalogirou@mail.ntua.gr, fkoutroulis@ihep.ac.cn}
\end{center}
\vskip 1.5 true cm
\centerline {\bf Abstract}
\vskip 3.0ex
\thintablerule
\vskip 2.0ex

We argue that a scalar field in de Sitter spacetime should feel explicit thermal effects associated with its curvature. Starting from the 
Bunch-Davies vacuum and a scalar field with small mass compared to the de Sitter curvature, we use the thermo-field dynamics formalism 
in order to expose these thermal effects. We compute the thermal Wightman function connecting two spacetime points and
from it, via the point-splitting regularization technique, the renormalized thermal energy-momentum tensor. We then examine how these 
corrections affect the de Sitter geometry by solving for the semi-classical backreaction and find that their sign depends on the initial conditions. In order to place our results in context, we compare them to the corresponding 2-loop quantum gravity correction to the cosmological constant derived in \cite{WoodardTsamis1996}.

\vskip 2.0ex
\thintablerule

\vskip-0.2cm
\newpage

\section{Introduction}

The study of de Sitter (dS) space-time has been one of the forefronts of early universe cosmology for nearly six decades now, mainly due 
to its usefulness in the various inflationary scenarios that attempt to model the first epoch of the universe's violent expansion. Since dS is a solution to the Einstein 
equations in the presence of a positive cosmological constant $\Lambda$, it provides a fitting background that scales the metric in time $t$ by a factor $a(t)$, 
as long as the conformally flat coordinate chart is used. 
de Sitter space in four dimensions being characterized by an $SO(1,4)$ isometry, an enhancement of the four-dimensional Poincare symmetry of flat space, 
has proven to be the ideal toy spacetime in our attempts for the understanding of quantum effects that involve matter in curved backgrounds. 
One of the prominent features of quantum fields in backgrounds without time translation symmetry is the lack of the notion of a global vacuum state.
This complicates the analysis of various physical phenomena with respect to flat space considerably, so it is only natural to first try to understand
the system's behaviour in a simple paradigm, which in our case will be that of a free, light scalar field in dS background.

Perhaps the most famous choice of a vacuum state in dS is the so called Euclidean vacuum, defined in \cite{Allen} as the one invariant under the full $SO(1,4)$ symmetry. 
Via a Bogolyubov Transformation (BT) with parameters $c = \cosh \a,d = e^{i \gamma} \sinh \a$, a whole family of other vacuum states can be constructed from the 
Euclidean vacuum, out of which only those with $\gamma = 0 $ enjoy the full dS isometry.
In any other case ($\gamma \neq 0 $), the resulting state will contain an infinite number of particles relative to the Euclidean vacuum. 
The Bunch Davies (BD) vacuum state $\ket{0,{\rm BD}}$ \cite{bunch1978} 
which was first discussed by Chernikov and Tagirov \cite{Tagirov1968}, is simply defined as the Euclidean vacuum whose modes have 
masses $m$ that satisfy $m^2 >0$, while the corresponding two point functions behave at short distances, as flat. 

Since the original work of Gibbons and Hawking \cite{GibbonsHawking} which provided strong indications that a time-like observer in dS should measure 
a thermal density of particles, there have been many attempts to calculate this particle production under different circumstances. Initially, an analog of the Unruh detector following a time-like curve  in a spacetime containing an event horizon is considered and the phenomenon of observer-dependent particles in curved backgrounds is discussed for the first time. 
Then \cite{Lapedes1978} argued that in the semi-classical context, the same thermal spectrum ought to appear independently of the choice of the initial `in' vacuum state.  
By describing the event horizon of the expanding patch of dS via similar static coordinates, the Hawking effect was reproduced by considering the response of the corresponding fields to the BD vacuum \cite{Higuchi1987}. 
In \cite{Mottola1985}, Mottola computed the decay rate of particle creation between an $\ket{\rm in }$ and an $\ket{\rm out }$ state which where conveniently 
parametrized so that the Euclidean vacuum rested in between them. In \cite{Mottola1986}, he discussed how quantum fluctuations in curved backgrounds result to a dissipative vaccuum that could potentially be responsible for the end of the inflationary period.  
One step further, \cite{Mottola2000} considered the adiabatic vacuum via the 
WKB approximation along with a time-dependent particle number arising from a time-dependent BT. There, the particle production is calculated numerically 
and is seemingly happening at the time of horizon exit for the massive case, in contrast to the massless case where restrictions from conformal symmetry hold. Interestingly, \cite{Prokopec2004} starting from the BD vacuum, used a BT in order to calculate the produced number density and associated it with a thermal equilibrium.  

A natural question is whether the produced particles should trigger through the Einstein equations a
backreaction to the original geometry. In order to formulate this effect, the vacuum expectation value (VEV) of the Energy-Momentum Tensor (EMT) 
of the produced spectrum should be calculated. 
Over the years, there have been various endeavors to calculate the renormalized EMT, each incorporating a different regularization procedure or approximation. Most commonly, the adiabatic regularization assumes a slow varying background and a WKB expansion of the modes while an improvement to this method was discussed in \cite{Anderson2000}.
On the other hand, Bunch and Davies \cite{bunch1977, bunch1978} using the point-splitting regularization technique, extracted the regulated Green's function of a massive free scalar field propagating in dS and from 
that the regulated VEV of the EMT. 
The thermal propagator and the resulting renormalized EMT for a conformally invariant field in an ultrastatic metric was first calculated in \cite{Page1982}.
Over the years, there has been evidence that indicates the instability of the BD state, or dS in general from both quantum and gravity origins. This was accomplished in \cite{Ford1985} by considering how interactions affect the renormalized EMT at firs order while \cite{Mottola1986_2} showed that the one-loop effective action breaks the $O(1,4)$ symmetry. On the gravity side, \cite{Antoniadis1986} showcased that the renormalized equation of motion of the graviton results into a vanishing scalar curvature.    
Moreover, in \cite{Mottola2014, Mottola2014_2} the propagation of a scalar field in dS was compared to the propagation of a charge in an electric field with the 
latter producing a known backreaction. It was concluded that the BD state is unstable under particle production which was argued again in  \cite{Mottola2018}.

Based on the previous discussion, the central quantity of interest for us is again the expectation value of the EMT, however through a different point of view.
Let us explain this.
The main assumption in \cite{bunch1978} 
is that both the state and the ladder operators are defined using the full dS isometry. As a consequence, while the resulting VEV -even though time-dependent- automatically satisfies the thermal 
conditions at the characteristic dS temperature $T_{\rm dS}$, it is seen to any co-moving observer at any given time as a zero-temperature VEV, in the sense that
\be\label{BDVEV}
\braket{0,{\rm BD} | \{\phi\phi\}_{{\rm BD}}   | 0,{\rm BD}} \sim u^* u + u u^* \braket{0,{\rm BD}| {\hat N}_\tmk^{\rm BD} | 0,{\rm BD}} + {\cdots}
\ee
where $u$ is a mode function, ${\hat N}_\tmk^{\rm BD}$ is the BD particle number operator and the dots indicate vanishing contributions.
Clearly, as it stands, the expectation value ${N}_\tmk^{\rm BD}\equiv \braket{0,{\rm BD}| {\hat N}_\tmk^{\rm BD} | 0,{\rm BD}} = 0$.
To put it simply, the BD VEV contains only implicitly the information about its thermal nature, by satisfying a certain periodicity (KMS) condition with periodicity parameter the $T_{\rm dS}$.
Here we extend the notion of the thermal VEV by observing that since the expanding spacetime produces a particle bath, the only way an observer can feel the 
corresponding effects at any time is by first thermally transforming the BD vacuum and then computing the expectation value. Then the VEV 
on the right hand side of \eq{BDVEV} becomes non-zero.
In addition we study how these thermal fluctuations affect the background by calculating the 
backreaction on the metric triggered by the thermal VEV of the EMT. 
In fact since matter necessarily breaks the full isometry to some extent, such an operation seems inevitable in models where a quantum field 
is coupled to the dS background. The thermal transformation of the BD state can be performed by means of the Thermo-Field Dynamcis (TFD) formalism \cite{Takahashi}.\footnote{For an extra 
motivation regarding the use of TFD, the interested reader could see \cite{FotisAntonis2}, where a momentum space analysis has been done. There it is 
argued that the breaking of the scale symmetry in the cosmic microwave background (CMB) is due to such thermal effects.} Then, by the use of the point-splitting regularization 
we follow analogous steps to \cite{bunch1978} and calculate the right hand side of the Einstein equations. The resulting metric corresponds
to the new geometry, a sort of deformation of dS, due to the backreaction to the thermal spectrum. To our knowledge, while there has been a considerable 
amount of interest in the backreaction problem due to quantum fluctuations \cite{Mottola2005, Nadal2008}, nobody has made use of real-time formalisms combined with the point splitting regulator. As we will see, 
this combination comes with certain advantages as regards our goal of obtaining the imprint of the intrinsic thermal effects on the geometry, resulting from non-local in time measurements. Furthermore, we point out that the steps presented here can be extended to cases where the thermal vacuum is not fixed at past infinity as long as one is interested in subhorizon modes or the overall mass parameter is small. 

This work is organized as follows. In \sect{Mottola-BD} we introduce the problem and build the TFD formalism which we will use in order to give the 
general form of the VEV of the thermal EMT. In addition, we give further arguments to motivate the computation that follows and explain where it differs from previous approaches. 
In \sect{Thermal_EMT} we compute the thermal VEV of the EMT at dS-temperature which we renormalize via the point-splitting technique and 
in \sect{thermal backreaction} we solve the Einstein equations. This leads to the perturbed metric that deviate the system away from the original 
dS metric due to the thermal effects, which we will interpret and compare to other known results. \sect{Conclusions} will contain our conclusions and a discussion 
for future work. In the Appendices we review the original derivation of the Bunch-Davies VEV and backreaction.

\section{Definition of the thermal Energy-Momentum Tensor} \label{Mottola-BD}

Our main goal is to compute the semi-classical backreaction to gravity from those thermal effects in dS space that develop when measurements are performed
at a time instant different from the time instant where the vacuum state has been defined. 
Semi-classical backreaction effects can be incorporated in general by the semi-classicaly generalized Einstein equations
\bad \label{EinEq.}
G_{\mu \nu} + \Lambda g_{\mu \nu}= 8 \pi G \braket{T_{\mu \nu}}\, ,
\ead
with $T_{\mu \nu}$ the EMT of the theory. In our case this right hand side will come from the action of a massive free scalar field, non-minimally coupled to gravity, in $d+1$ dimensions:
\bad \label{ActionS}
{\cal S} = \frac{1}{2} \int d \t \ d^{d} x \ \sqrt{-g} \left\{g^{\mu \nu} \partial_\mu \phi \partial_\nu \phi  + \mu^2 \phi^2 \right\}, \qquad \mu^2 = m^2 + \xi {\cal R} \ .
\ead
We will restrict ourselves to the expanding Poincare patch of dS spacetime via the conformally flat metric
\bad
ds^2 = a^2(\t) (-d \t^2 + dx^2)\, ,
\ead
where $\t$ is the conformal time, taking values in $\tau\in (-\infty, 0)$. 
The time instant $\tau \rightarrow -\infty$ is the beginning of this spacetime and it is where (or perhaps better when) the 
$\ket{0,{\rm BD}}$ vacuum is defined. On the other hand $\tau \rightarrow 0$ signals for us the end of the spacetime, called the Horizon.
It is on the Horizon where we will place our observer measuring the energy and pressure densities. 
Of particular interest is the dS metric determined by $a(\tau) = - \frac{1}{H\tau}$. For a general FRW metric the Hubble parameter $H \equiv \frac{a'}{a^2}$ (in conformal coordinates) is a function of time,
however for dS it is a constant, related to the scalar curvature as ${\cal R} = d(d+1) H^2$. Sometimes this constant is exchanged for the cosmological constant parameter
via $\Lambda =  \frac{d(d-1)}{2}H^2$. 
The Hamiltonian of the theory is
\bad
\hat H (\t) = -\frac{1}{2} a^{d-1} \int d^d \bm x \left\{ (\partial_0 \phi)^2 + \delta^{ij} \partial_i \phi \partial_j \phi + a^2 \mu^2 \phi^2 \right\} 
\ead
and the classical EMT is defined as the functional derivative of the above action with respect to the metric which leads to:
\bad \label{Tmn}
T_{\mu \nu} &\equiv \frac{2}{\sqrt{-g}} \frac{\delta {\cal S}}{\delta g^{\mu \nu}} \\
& = \partial_\mu \phi \partial_\nu \phi - \frac{1}{2} g_{\mu \nu} g^{\kappa \lambda} \partial_\kappa \phi \partial_\lambda \phi  - 
\frac{1}{2}  m^2 g_{\mu \nu} \phi^2 - \xi \nabla_\mu \nabla_\nu \phi^2 + \xi g_{\mu \nu} g^{\kappa \lambda} \nabla_\kappa \nabla_\lambda \phi^2 + \xi G_{\mu \nu} \phi^2\, ,
\ead
with the last three terms originating from the non-minimal $\xi$-coupling term in the action.  

In order to compute the quantum VEV of the above expression, one first expands the scalar field $\phi$ as
\bad \label{scalarfield}
\phi(\tau, \tx) = \frac{1}{(2 \pi)^{d/2}}\int d^d \tk \ \left[ u^*_\tmk(\t) a^-_\tk + u_\tmk (\t) a^+_{-\tk} \right] e^{i \tk \tx}\, .
\ead
The $u_\tmk$ and $u^*_\tmk$ are the mode functions that form a basis of solutions to the classical equation of motion (eom):
\bad \label{Covarianteom}
u''_\tmk + (d-1)\frac{a'}{a} u'_\tmk + \left( \tk^2 + a^2 \mu^2 \right)u_\tmk = 0\, ,
\ead
where the prime denotes differentiation with respect to the conformal time $\t$.
The resulting quantum Hamiltonian is
\begin{subequations}
\bad
\hat H (\t) = -\frac{ a^{d-1}(\t)}{2 }  \int d \bm k \left\{ \left(a^-_\tk a^+_\tk + a^+_\tk a^-_\tk \right) E_\tmk(\t) + a^+_\tk a^+_{-\tk} F_\tmk(\t) + a^-_\tk a^-_{-\tk} F^*_\tmk (\t) \right\} \, ,
\ead
with
\begin{gather} \label{EFProkopec}
E_\tmk(\t) =   \left| {u'}_\tmk \right|^2 + \o^2_\tmk(\t)  \left| {u}_\tmk \right|^2 \\
F_\tmk(\t) =  \left( {u'}_\tmk \right)^2 + \o^2_\tmk(\t) \left( {u}_\tmk \right)^2 
\end{gather}
and
\bad \label{omega}
\o^2_\tmk (\t) = \tk^2 + a^{2}(\t) \mu^2
\ead
\end{subequations}
is the corresponding angular frequency. The most general solution to \eq{Covarianteom} in dS space
can be written in terms of the Bessel functions $J_\nu, Y_\nu$ as
\bad \label{uk}
u_\tmk = a^{-\frac{d-1}{2}} \sqrt{\tmk |\tau|} \left[ A_\tmk J_\nu (\tmk |\tau|) + B_\tmk Y_\nu (\tmk |\tau|) \right], \qquad  \nu = \sqrt{\frac{d^2}{4} - \frac{\mu^2}{H^2}}\, ,
\ead
where the $A_\tmk, B_\tmk$ coefficients satisfy $A_\tmk B^*_\tmk - A^*_\tmk B_\tmk = \frac{i \pi }{2 \tmk}$ so that the Wronskian 
satisfies $a^{d-1} W\left(u_\tmk, u^*_\tmk \right) = i $. It is then obvious that there is a freedom of choice for $A_\tmk, B_\tmk$ with the most famous being 
\bad
A^{\rm BD}_\tmk = \frac{1}{2}\sqrt{\frac{\pi}{ \tmk}}, \qquad B^{\rm BD}_\tmk = - i A^{\rm BD}_\tmk 
\ead
so that the corresponding solution
\bad \label{ukBD}
u_{\tmk;\rm BD} = \frac{\pi^{1/2} H^{\frac{d-1}{2}} }{2}|\tau|^{d/2} H^{(2)}_\nu (\tmk |\tau|)
\ead
defines the so called BD mode function.

Quantization proceeds by imposing the creation and annihilation operators $a^-_\tk, a^+_\tk$ to satisfy the commutation relations
\be
[a^-_\tk, a^+_\tp] = \delta^{(3)} (\tk - \tp ) \, , \hskip 1cm [a^-_\tk, a^-_\tp] = [a^+_\tk, a^+_\tp] = 0\, .
\ee
These can be used to define a vacuum state by
\be \label{DefVacuum}
a^-_\tk \ket{0} = 0. 
\ee
Since there is no unique way to define a vacuum, there is no unique basis of creation/annihilation operators to expand the field $\phi$. 
As a result, we must generalize the expansion \eq{scalarfield} by adding the label $i$ that specifies the vacuum state which the 
operator $a^-_{\tk;i}$ annihilates: 
\be
a^-_{\tk;i} \ket{0,i} = 0\, .
\ee
The field is then written as
\bad \label{scalarfieldi}
\phi_i(\tau, \tx) = \frac{1}{(2 \pi)^{d/2}}\int d^d \tk \ \left[ u^*_{\tmk;i}(\t) a^-_{\tk;i} + u_{\tmk;i} a^+_{-\tk;i} \right] e^{i \tk \tx}\, 
\ead
with the operators satisfying
\be
[a^-_{\tk;i}, a^+_{\tp; j}] = \delta^{(3)} (\tk - \tp ) \ \delta_{ij}\, , \hskip 1cm [a^-_{\tk,i}, a^-_{\tp;j}] = [a^+_{\tk,i}, a^+_{\tp;j}] =  0\, .
\ee
The BT is defined as the linear transformation which connects the two bases:
\bad \label{BT}
a^-_{\tk;i} = c^*_{\tmk;ij} a^-_{\tk ; j} - d^*_{\tmk; ij} a^+_{-\tk; j}, \qquad a^+_{-\tk;i} = c_{\tmk;ij} a^+_{-\tk;j} - d_{\tmk;ij} a^-_{\tk ; j}
\ead 
where the BT coefficients may depend on time \cite{Mottola2000} but always have to satisfy the normalization condition $|c_{\tmk;ij}|^2 - | d_{\tmk; ij}|^2 = 1$. 
The mode functions must satisfy an inverse BT
\bad \label{BTu}
u_{\tmk; i} = c^*_{\tmk;ij} u_{\tmk; j} + d^*_{\tmk;ij} u^*_{\tmk; j}\, ,
\ead
so that the quantum field \eq{scalarfieldi} remains invariant under the change of basis. 
This implies that the index $i$ on the field $\phi_i$ is just a label to remind of the vacuum state that the creation/annihilation operators 
and the corresponding mode functions see inside the field. The above formulae are usually interpreted as indication of particle production between 
different vacuum states, since positive-frequency modes in one basis are linear transformed into positive and negative modes in another. 

In general, we define for any quantum, composite operator $A_i$ constructed from the field $\phi_i$ and its derivatives, the VEV
\be \label{BTA}
\braket{A_i}_j \equiv \braket{0;j| A_i  | 0;j}\, .
\ee
Clearly, when $i=j$ we have mostly the quantities reviewed in the first part of the Introduction. When $i\ne j$ we obtain much less investigated observables.
Our intuition regarding these latter observables is that they must be necessarily sensitive to the intrinsic dS temperature, just as an observer 
sitting at, say the north pole of a sphere sees the curvature when looking at any other point, in contrast to when he looks at the north pole itself,
in which case he sees no curvature. In dS, the sphere's curvature translates to temperature while the different points on the sphere 
into different time instances that is, different vacua.

The flagship $i=j$ calculation is that of Bunch and Davies \cite{bunch1978}, who considered a scalar field propagating in 4d dS spacetime 
and used \eq{ukBD} for d=3  and the corresponding state $\ket{0,{\rm BD}}$ to compute
\bad\label{GBDdiag}
G_{\rm BD} (x_1 - x_2)  = \braket{0, {\rm BD} | \phi_{\rm BD}(x_1) \phi_{\rm BD}(x_2) | 0, {\rm BD} }  \equiv \braket{\phi_{\rm BD}(x_1) \phi_{\rm BD}(x_2)}_{\rm BD}\, ,
\ead
which is the Wightman function of the field in the BD state. Then,
\bad \label{GBD_0}
G_{\rm BD} (x_1 - x_2) = \frac{1}{(2 \pi)^{d}} \int d^3 k \ u^*_{\tmk; \rm BD} (\t_1) \ u_{\tmk ; \rm BD} (\t_2) \ e^{i \tk \Delta \tx}\, ,
\ead
where $\t_1, \t_2$ are the (conformal) time coordinates corresponding to $x_1,x_2$ and $\Delta \tx = \tx_1 - \tx_2$ is the spatial distance between the two points. 
Defining the dimensionless parameter $\chi = 1     +  \frac{\Delta s^2}{2 \tau_1 \tau_2}$, where $\Delta s^2 = (\t_1 - \t_2)^2 - \Delta x^2$, in four dimensions ($d=3$) they found
\bad \label{GBD}
G_{\rm BD} (x_1 - x_2) = \frac{H^2}{16 \pi^2}  \Gamma \left(\frac{3}{2} + \nu\right) \Gamma \left(\frac{3}{2}- \nu\right)  
F \left( \frac{3}{2} + \nu,\frac{3}{2} - \nu,2;\frac{1}{2} (\chi + 1)\right)\, .
\ead
Notably, the above expression is invariant under the imaginary cosmic time translation $t \rightarrow t + i \beta/2$. 
Recall that the cosmic time is defined by $d\t = a^{-1}(t) d t$ so that in terms of the conformal time, this translation becomes 
the phase transformation $\tau \rightarrow e^{-i H \beta/2} \tau$ which becomes trivial for 
\be
\beta \equiv \frac{1}{T_{\rm dS}} = \frac{2 \pi}{ H} \, .
\ee
This means that at exactly the dS temperature $T_{\rm dS}$, the BD result is periodic under imaginary cosmic time translations suggesting 
that \eq{GBD} can be considered as a thermal Wightman function since it satisfies the so called KMS condition.   

The expression \eq{GBD} contains UV divergences arising from the hypergeometric function $F(a,b,c;z)$ when $\chi =1$, that need to be somehow subtracted. 
As demonstrated in detail in Appendix \ref{Point-splitting}, \cite{bunch1978} dealt with the above divergences using the point-splitting regularization, 
which is based on the idea that the two distinct space-time points $x_1 = (\t_1, \tx_1)$ and $x_2 = (\t_2, \tx_2)$ are endpoints of the 
same geodesic line whose length $2\epsilon$ plays the role of its affine parameter (see \fig{Fig_Parallel_Transport}). 
Since the two points are separated by a proper distance equal to $2 \epsilon$, we can regulate the expression by expanding 
the hypergeometric function around $\epsilon \rightarrow 0$ and discard all terms of order $\epsilon^{-1}$ or lower. 
For completeness, in Appendix \ref{Point-splitting}, we repeat the calculation of the BD EMT, given by the expression
\bad
\braket{ T^{\rm BD}_{\mu \nu}}_{\rm BD} &= \braket{ \partial_\mu \phi_{\rm BD} \partial_\nu \phi_{\rm BD} }_{\rm BD}   - 
\frac{1}{2} g_{\mu \nu} g^{\kappa \lambda} \braket{ \partial_\kappa \phi_{\rm BD} \partial_\lambda \phi_{\rm BD}}_{\rm BD}  
- \frac{1}{2}  m^2 g_{\mu \nu} \braket{ \phi^2_{\rm BD}  }_{\rm BD} \\
& \qquad - \xi \braket{\nabla_\mu \nabla_\nu \phi^2_{\rm BD}}_{\rm BD} + \xi g_{\mu \nu} g^{\kappa \lambda} 
\braket{\nabla_\kappa \nabla_\lambda \phi^2_{\rm BD}}_{\rm BD} + \xi G_{\mu \nu} \braket{\phi^2_{\rm BD}}_{\rm BD},
\ead
by first determining the Wightman function \eq{GBD_0}, act with the derivatives, expand in infinitesimal point splitting and in the small parameter 
\be
\eta\equiv \frac{{\cal R}}{m^2} = 48\pi^2 \frac{T_{\rm dS}^2}{m^2}
\ee
and finally regulate by removing the divergences. Then, the renormalized energy-momentum tensor turns out to be, up to order $\eta^2$, equal to
\bea \label{TmnBD}
\braket{ T^{\rm BD}_{\mu \nu}}_{\rm BD} &=& -\frac{1}{4} m^2 g_{\mu \nu} G_{\rm BD} \Big|_{\epsilon \rightarrow 0} \nonumber\\
 &= &-\frac{g_{\m \n}}{64 \pi^2} m^4 \biggl\{ \left[ 1 + \eta 
\left( \xi - \frac{1}{6} \right) \right]  \left[ \psi \left( \frac{3}{2} + \nu \right) + \psi \left( \frac{3}{2} - \nu \right) + \ln\left(\frac{\eta}{12}\right) \right] \nonumber\\
&-&  \left( \xi - \frac{1}{9}  \right) \eta  -  \eta^2 \left[\frac{1}{2} \left( \xi - \frac{1}{6} \right)^2  - \frac{1}{2160 } \right]\biggr\} \, .
\eea
Here, $\psi(z)$ is the polygamma function. The above result corresponds to the energy-momentum that an observer would measure in his own vacuum state.

However, since dS spacetime has a Horizon, any geodesic observer using an Unruh detector detects a thermal bath of particles at a temperature 
$T_{\rm dS}$. We suspect that this should be connected with the continuous horizon crossing of the modes \eq{uk} at each time-instance, 
which corresponds to a loss of information from the geodesic observer's point of view. 
The foresight that near the end of inflation the perceived system should be characterized by a thermal equilibrium of non-zero particle number, 
motivates us to build a mechanism in order to describe the arising effects due to the implicit temperature. This leads us to use the 
TFD formalism \cite{Takahashi} in order to rotate the BD state so that we can define the corresponding VEVs and extract the hidden thermal corrections 
on basic quantities such as the BD Wightman function and the EMT tensor that is derived from it. Hence, what we are really interested in here is the $i\ne j$ quantity
\bad \label{inoutG}
G_{\rm \beta} (x_1 - x_2) = \braket{0, \b| \phi_{\rm BD} (x_1)  \phi_{\rm BD} (x_2)   | 0, \b} \equiv \braket{\phi_{\rm BD} (x_1)  \phi_{\rm BD} (x_2)}_{\b}
\ead
and the resulting thermal VEV $\braket{T_{\m\n}}_\beta$. Here $\ket{0 ,\b}$ is the thermal state which contains a particle number equivalent to 
the Bose-Einstein number density $n_{{\rm B},\tmk}$ and is obtained via a transformation of the original BD vacuum, as we will see below. Note that in \cite{Maldacena2015} similar VEVs to \eq{inoutG} have been calculated. However, the latter involve composite operators defined in CFTs of an AdS boundary, they grow at late times exponentially and provide a specific bound on their Lyapunov exponent. Therefore this analysis does not overlap with the current work.

Choosing to describe the thermal effects in the formalism of TFD comes with two important advantages. Firstly, the resulting thermal state 
will retain its pure properties letting us avoid the complicated case of a mixed thermal state. Thus \eq{inoutG}
can be calculated directly, by expanding the scalar field \eq{scalarfield} in terms of the BD mode functions and consider its VEV with respect 
to the thermal vacuum. Then, we choose to deal with the arising UV divergences via point-splitting, extending the work of BD while also distancing 
at this point our approach from others. An extra motivation supporting point-splitting is that it provides a nice framework where a local observable  
like the EMT can be seen as a limit of a non-local function of the proper distance. Temperature on the other hand as explained is a non-local characteristic, 
hence after having obtained the renormalized thermal EMT by removing the point splitting, the interesting question whether the thermal effects persist, arises.
An additional advantage of TFD is that it does not force us to sacrifice our time-coordinate via an analytic continuation as the imaginary-time formalism does. 
At the same time, we use a time-dependent BT \eq{BT} between our observers with the coefficients $c_{\tmk;ij}(\t), d_{\tmk ; ij}(\t)$ 
being functions of the conformal time. Then the TFD transformation can be seen as a BT that introduces naturally the intrinsic temperature in the system. 
Let us briefly explain the the central idea of TFD and adjust it to our framework. 

We double the degrees of freedom by considering a copy of 
the initial Hilbert space. The resulting extended Hilbert space contains the tensor product states
\bad
\ket{m,\tilde n} = \ket{m} \otimes \ket{\tilde n}\, .
\ead
Accordingly, the field needs to be generalized as
\bad \label{TFDfield}
\Phi(\t,\tx) = \begin{pmatrix}
\phi(\t,\tx) \\
\tilde \phi(\t,\tx) 
\end{pmatrix}\, ,
\ead
where
\bad \label{tildescalarfield}
\tilde \phi(\tau, \tx) = \frac{1}{(2 \pi)^{d/2}}\int d^d \tk \ \left[ u^*_{\tmk}(\t) \tilde a^+_{\tk} + u_{\tmk }(\t) \tilde a^-_{-\tk } \right] e^{i \tk \tx}
\ead
is the tilded field expanded in terms of its tilded ladder operators. The total action of the doubled system is then defined as:
\bad \label{Stot}
{\cal S}_{\rm tot} (\t, \tx) &= {\cal S} (\t, \tx) - \tilde {\cal S}(\t, \tx) \\
&= \frac{1}{2} \int d \t \ d^{d} x \ \sqrt{-g} \left\{g^{\mu \nu} \left( \partial_\mu \phi \partial_\nu \phi -  \partial_\mu \tilde \phi \partial_\nu \tilde \phi \right)  + \mu^2 \left(\phi^2 - \tilde \phi^2 \right) \right\} \\
&= \frac{1}{2} \int d \t \ d^{d} x \ \sqrt{-g} \left\{g^{\mu \nu}  \partial_\mu \Phi^T \sigma_3 \partial_\nu \Phi + \mu^2 \Phi \sigma_3 \Phi  \right\}\, ,
\ead
where we used the Pauli matrix $\sigma_3 = {\rm diag}(1,-1)$ in order to write the action in a more condensed form.
Furthermore, we introduce the matrix transformation 
\bad \label{TFDa}
\begin{pmatrix}
a^-_\tk  \\
\tilde a^+_\tk 
\end{pmatrix}
=
\begin{pmatrix}
\cosh \theta_\tmk (\beta) & \sinh \theta_\tmk (\beta) \\
\sinh \theta_\tmk (\beta) & \cosh \theta_\tmk (\beta)
\end{pmatrix}
\begin{pmatrix}
a^-_{\tk;\b}  \\
\tilde a^+_{\tk;\b} 
\end{pmatrix}
=
U_\b(-\theta_\tk) 
\begin{pmatrix}
a^-_{\tk;\b}  \\
\tilde a^+_{\tk;\b} 
\end{pmatrix}
\ead
from which we can define the thermal pure state from its zero-temperature vacuum counterpart, as
\bad
\ket{0, \beta} = {\cal U}_\b  \ket{0, \tilde 0}, \qquad {\cal U}_\b  = e^{- \int d^d k  \ \theta_\tmk (\beta) \left(\a^-_\tk \tilde \a^-_\tk - \a^+_\tk \tilde \a^+_\tk \right) },
\ead
where the angle $\theta_\tmk ( \beta)$ is a function of the inverse temperature $\b = 1/T$ so that 
\bad \label{TFDnB}
\sinh^2 \theta_\tmk (\beta) = \frac{e^{-\beta \o_\tmk}}{1 - e^{- \beta \o_\tmk}} \equiv n_{B;\tmk}\, 
\ead
is equal to the Bose-Einstein number density. 
Notice that due to the non-zero curvature, the number density contains the time-dependent frequency \eq{omega} and as a result the TFD coefficients
obtain an explicit time-dependence indicating that different times are characterized by different number of particles. The zero-temperature ladder operators 
on the left hand side of \eq{TFDa} are still time-independent, while those on the right hand size are clearly not. 

The proper thermal VEV of an observable $A$ constructed from $\Phi$ in \eq{TFDfield} and its derivatives is then defined as:
\bad
\braket{A}_\beta \equiv \braket{0, \beta | A | 0 , \beta},
\ead
where $A$ is understood to be expanded in terms of the zero-temperature ladder operators. Then by using \eq{TFDa}, one can write the observable 
$A$ in the thermal basis and proceed by computing the action of the corresponding operators on the thermal vacuum exactly as one would do in the non-thermal case. 
This process resembles the calculation of \eq{BTA} via BT, with $\ket{0;j} \equiv \ket{0;\beta}$, indicating that TFD is just a generalization of the latter. 
It should be noted that although both BT and TFD transformations follow similar logic, they are not exactly equivalent. 
The BT keeps the field itself (therefore also the action) invariant via the inverse transformation \eq{BTu}. 
In contrast, the TFD starts from the rotation \eq{TFDa} which also transforms of course the (Fourier transform of the) field 
\eq{TFDfield}, in such a way that the matrix $U_\beta$ cancels out in the action \eq{Stot}.
 
After these clarifications, we can define our thermal EMT as
\bad \label{Tmnbtot}
\braket{T_{\mu \nu}}_\beta \equiv \braket{0, \beta | T_{\mu \nu} | 0 , \beta}\, ,
\ead
where 
\bad
T_{\mu \nu} &= (\partial_\mu \Phi)^{T} \sigma_3 \partial_\nu \Phi - \frac{1}{2} g_{\mu \nu} \left(g^{\kappa \lambda} 
(\partial_\kappa \Phi)^T \sigma_3 \partial_\lambda \Phi  + m^2 \Phi^T \sigma_3 \Phi \right)  \\
& \qquad  - \xi \left\{  \nabla_\mu \nabla_\nu \Phi^T \sigma_3 \Phi -  g_{\mu \nu} g^{\kappa \lambda} \nabla_\kappa \nabla_\lambda \Phi^T \sigma_3 \Phi -  
G_{\mu \nu} \Phi^T \sigma_3 \Phi \right\}
\ead
is the EMT corresponding to the action \eq{Stot}, with the fields $\phi$ and $\tilde \phi$ in $\Phi$ expanded with the zero-temperature 
ladder operators. 

In the point splitting spirit, we need to know first the thermal Wightman functions
\begin{subequations}
\bad
G_{\b} (x_1-x_2) =  \braket{ \phi(x_1) \phi(x_2)}_\beta 
\ead
\bad
\tilde G_{\b} (x_1-x_2) =  \braket{ \tilde \phi(x_1) \tilde \phi(x_2)}_\beta 
\ead
\end{subequations}
Using the TFD transformation \eq{TFDa} along with \eq{scalarfield} and \eq{tildescalarfield} we obtain
\begin{subequations}
\begin{align} \label{G1b}
G_{\b} (x_1-x_2) &= \frac{1}{(2 \pi)^d} \int d^d k \left[ u^*_\tmk (\t_1) u_\tmk (\t_2) \cosh^2 \theta_\tmk (\beta)  + u^*_\tmk (\t_2) u_\tmk (\t_1) \sinh^2 \theta_\tmk (\beta) \right] e^{i \tk (\tx_1 - \tx_2)} 
\end{align}
and
\bad\label{G1btilde}
\tilde G_{\b} (x_1-x_2) = \frac{1}{(2 \pi)^d} \int d^d k \left[ u^*_\tmk (\t_1) u_\tmk (\t_2) \sinh^2 \theta_\tmk (\beta)  + u^*_\tmk (\t_2) u_\tmk (\t_1) \cosh^2 \theta_\tmk (\beta) \right] e^{i \tk (\tx_1 - \tx_2)}\, .
\ead
\end{subequations}
The thermal functions $\sinh\theta_\tmk (\beta)$ and $\cosh\theta_\tmk (\beta)$ depend implicitly on the time instance $\tau_{\rm in}$, where the transformation is implemented. Now the time $\tau_{\rm in}$ is different from $\tau_{1,2}$ which are two arbitrary time instances in the space-time's interior. Then inspecting the above expressions shows that the contribution of the untilded and tilded sectors to the VEV are equal, yielding a vanishing result due to the relative minus sign between the two sectors. 
This is actually no paradox, as can be seen by recalling that at the Wightman function level the Thermo-Field is equivalent to the Schwinger-Keldysh (SK) representation, which is a path integral along the closed time contour $\tau_{\rm in}\to \tau_{\rm out}\to \tau_{\rm in}$. As such, it naturally yields a vanishing energy-momentum change. This argument also shows that for an 'in to out' process it is sufficient to take into account only half of the contour, that is only one of the sectors in the TFD formalism. Note that the same corrections could be obtained via the SK formalism as shown in \cite{SemenoffWeiss}. The relevant quantity for us is the physical EMT which in the language of TFD corresponds to the untilded part of the VEV otherwise there is a redundant double counting. Hence, we define the physical EMT
with $T_{\m \n}$ given by \eq{Tmn}, the EMT of the untilded sector. 

The point-splitting regularization requires to write the quantity as the double limit of two points $x_1$ and $x_2$ that are 
parallel transported along the spacetime geodesic towards the point $x$, where the EMT is defined. Consequently, we need to break down 
each term in \eq{Tmn} into a symmetric expression containing derivatives that depend on $x_1$ or $x_2$ and vice versa. Concretely:
\begin{align} \label{Tmnb2}
\braket{T_{\mu \nu}(x)}_\b &=  \lim_{\substack{x_1 \rightarrow x \\ x_2  \rightarrow x}} \frac{1}{2} (1 - 2 \xi) 
\Bigl(\nabla_{\mu,1} \nabla_{\nu,2} G_{\b} (x_1 - x_2) + \nabla_{\mu,2} \nabla_{\nu,1} G_{\b} (x_1 - x_2) \Bigr) \nn \\
& \qquad + \left(- \frac{1}{2} + 2 \xi \right) g_{\m \n} \nabla_{\kappa,1} \nabla^{\kappa,2} G_{\b} (x_1 - x_2) \nn \\
& \qquad - \xi  \Bigl(\nabla_{\mu,1} \nabla_{\nu,1} G_{\b} (x_1 - x_2) + \nabla_{\mu,2} \nabla_{\nu,2} G_{\b} (x_1 - x_2) \Bigr) \nn \\
& \qquad + \xi g_{\mu \nu}  \Bigl(\nabla_{\kappa,1} \nabla^{\kappa,1} G_{\b} (x_1 - x_2) +\nabla_{\kappa,2} \nabla^{\kappa,2} 
G_{\b} (x_1 - x_2) \Bigr)  \nn \\
& \qquad - \frac{1}{2} m^2 g_{\mu \nu} G_{\b} (x_1 - x_2) + \xi G_{\mu \nu} G_{\b} (x_1 - x_2) \, ,
\end{align}
where the subscripts $1,2$ keep track of the point on which the covariant derivative is defined. From \eq{G1b}, the thermal Wightman function can be then written as
\bad \label{G1bsplit}
G_{\b} (x_1-x_2) &= G (x_1-x_2) + \frac{1}{(2 \pi)^d} \int d^d k \left[ u^*_\tmk (\t_1) u_\tmk (\t_2)  + u^*_\tmk (\t_2) u_\tmk (\t_1)  \right]  \sinh^2 \theta_\tmk (\beta)  e^{i \tk (\tx_1 - \tx_2)}\, ,
\ead 
with $G(x_1 - x_2)$ being the zero-temperature function. Consequently, the thermal VEV of the EMT \eq{Tmnb2} will 
break into a non-thermal part and its thermal correction, enabling us to separate our calculation into two fragments. 

Actually, this is not the first time anyone has considered such an observable, so for completeness let us review the most relevant previous 
work. In his original paper \cite{Mottola1985}, 
Mottola calculated the probability that particles of any mode are created between some `in' and `out' states, which are defined as asymptotical 
positive and negative negative frequency states in the far past and future of the global dS.  Then, it was argued that the square of the transition 
amplitude $|\braket{\rm out|in}|^2$ decayed exponentially signaling a non-trivial particle production between the two states. Following, 
he computed the deviation of the vacuum energy via the difference:
\bad
\braket{  T^{\rm in}_{\m \nu}  }_{\rm in} - \braket{ T^{\rm out}_{\m \nu} }_{\rm out}  \, ,
\ead
which results into a decrease of the vacuum energy due to particle creation, which vanishes in the conformal limit. Afterwards 
in \cite{Mottola2005, Mottola2014, Mottola2014_2} time-dependent BT where used along with the adiabatic approximation in order to compute
\bad \label{Troutin}
\braket{ T^{\rm in}_{\mu \nu} }_{\rm out} \equiv  {\rm Tr} \left(\rho_{\rm out} T^{\rm in}_{\mu\nu}\right)\, ,
\ead
where the out state was considered as thermal, mixed and characterized by the mean particle number 
\bad \label{Nk}
N_\tk = {\rm Tr} \left(\rho_{\rm out} a^+_{\tk ; \rm out} a^-_{\tk ; \rm out}\right) = \sum_i \rho_i \braket{{\rm out}_i|a^+_{\tk ; \rm out} a^-_{\tk ; \rm out}|{\rm out}_i}  \ . 
\ead
Starting with \eq{Troutin}, by expanding the fields with respect to the BD modes \eq{ukBD}, the corresponding ladder operators need to be transformed 
via \eq{BT} into the out basis. For example, for $\xi = 0$ this produces a non-diagonal form for the 00 component of the EMT:
\bad \label{MottolasEMT}
\braket{ T^{\rm BD}_{0 0} }_{\rm out}  =\braket{ T^{\rm BD}_{00} }_{\rm BD} +  \frac{1}{8 \pi^3} \int d^3 \tk & \Bigl[- (1 + N_\tmk) {\rm Re} \Bigl(c_{\tmk;{\rm BD ,out}} d_{\tmk;{\rm BD ,out}} F_{\tmk; {\rm BD}} \Bigr)   \\
& \qquad + \Bigl(N_\tmk + |d_{\tmk;{\rm BD ,out}}|^2 ( 1 + 2 N_\tmk)  E_{\tmk; {\rm BD}} \Bigr) \Bigr] \ .
\ead
Then, a qualitative analysis was performed for the the corresponding semi-classical backreaction to the dS geometry. It concluded that there 
should be perturbations away from the BD state which correspond to large enough energy densities that alter the dS background. 

We can directly compare \eq{MottolasEMT} with \eq{Tmnb2} obtained via TFD once we substitute \eq{G1bsplit}. By nature, since the correction 
to the latter are proportional to $E_{\tmk; {\rm BD}}$ from \eq{EFProkopec} times the TFD coefficient $\sinh^2 \theta_\tmk (\b)$, it is straightforward to see that the relation:
\bad
 |d|^2_{\tmk; {\rm BD, out}} = \frac{\sinh^2 \theta_\tmk (\b) - N_\tmk }{1 + 2 N_\tmk} \geq 0
\ead
needs to be satisfied. Note that for $\rho_{\rm out} = \ket{0,{\rm BD}} \bra{0,{\rm BD}}$, the particle number $N_\tmk = 0$ meaning that $|d|^2_{\tmk; {\rm BD, out}}  = \sinh^2 \theta_\tmk (\b)$ which indicates that TFD are basically a BT that keeps the Hamiltonian diagonal. According to our interpretation since $\sinh^2 \theta_\tmk (\b) = n_{{\rm B}, \tmk}$, the above condition showcases that the 
total number of particles that the out state can contain, has as an upper bound the Bose-Einstein number density, which gives $|d|^2_{\tmk; {\rm BD, out}} = 0$ 
when the equality is satisfied. This results into a trivial BT which also causes the vanishing of the $u^*_{\tmk;{\rm BD}} u_{\tmk;{\rm BD}}$ terms 
in \eq{MottolasEMT}. Hence, the starting point of this current paper can be seen as an extension of \cite{Mottola2014_2} with the extra assumption 
that the Hamiltonian of the thermal state is diagonal. Exactly this distinction along with the use of the BD mode functions lets us then perform 
a theoretical analysis where the thermal corrections arise naturally from the super-horizon modes in Poincare coordinates.

\section{The renormalized thermal Energy-Momentum Tensor} \label{Thermal_EMT}

In this section, we compute the corrections to the Bunch-Davies EMT due to the appearance of a thermal bath at equilibrium. 
One first needs to calculate the thermal Wightman function given in \eq{G1bsplit}, while the calculation 
of the zero-temperature counterpart, is presented in App. \ref{Point-splitting}. Once, we have the full thermal Wightman function, 
we can make use of \eq{Tmnb2} in order to calculate the bare thermal EMT. Here, we will focus on the second term which for $d=3$ is:
\bad \label{dG1}
\delta G(x_1 - x_2) = \frac{1}{(2 \pi)^3} \int d^3  k \left[ u^*_\tmk (\tau_1) u_\tmk (\tau_2) + u^*_\tmk (\tau_2) u_\tmk (\tau_1) \right] \frac{1}{e^{\b \o_\tmk} - 1} e^{i \bm k \Delta \tx }\, ,
\ead  
where we substituted the TFD parameter:
\bad
\sinh^2 \theta_\tmk (\beta)  \equiv \frac{1}{e^{\b \o_\tmk} - 1}
\ead
and defined $\Delta \tx = \tx_1 - \tx_2$. Then, using the BD mode-functions \eq{ukBD}, polar-coordinates and the integral representation 
of the product of two Hankel functions \eq{ProdHankelRep}, we obtain
\bad \label{dG2}
\delta G(x_1 - x_2) = \frac{H^2}{4 \pi^3 |\Delta x|} |\t_1 \t_2 |^{3/2} \left[I(x_1, x_2) + I(x_2, x_1)  \right] 
\ead
with $I(\t_1, \t_2)$ a product of three integrals:
\bad \label{Iint}
I(x_1, x_2) =  \int^\infty_{-\infty} dT e^{-2 \nu T} \int^\infty_0 dv v^{-1} e^{-\frac{1}{2} v} \  {\rm Im} \left\{ \int^\infty_0 d \tmk \  \frac{\tmk}{e^{\b \o_\tmk} -1} e^{-\tmk^2 A + i \tmk \Delta x } \right\}\, ,
\ead
where 
\bad
A(\t_1,\t_2,v,T) = \frac{1}{v} \left[|\t_1 \t_2| \cosh 2T - \frac{1}{2} (\t_1^2 + \t_2^2 ) \right]
\ead
is a function of the two time-coordinates and the integration parameters $v,T$. Note that, even though we assume that $\t_1,\t_2$ 
are such that keep $A$ positive for all possible values of $v,T$ in order for the $\tmk$-integral to converge, we will deal with 
any persisting divergences later on.\footnote{It is straightforward to see that the condition $A>0$ is naturally satisfied in the limit $\t_1 \rightarrow \t_2$.}

We have reached a point where it is seems hard to find analytic solutions to the above integrals without making any additional assumptions or approximations. 
This is the reason why most previous works turned their attentions at this point to numerical calculations. However, by preparing the thermal vacuum at $\t_{\rm in} \rightarrow - \infty$ where the in frequency \eq{omega} satifies:
\bad
\omega_\tmk (\t_{\rm in})  \simeq \tmk 
\ead
and taking advantage from the fact that in the Heisenberg picture states do not evolve, we can make use of the above approximation in \eq{Iint}. This hypothesis can be further justified since we are concerned with how an observer at later times perceives the in state. Nevertheless, there are physically interesting cases for which this approximation holds for any $\t_{\rm in} \neq 0$ placed as long as $a(\t_{\rm in}) <1$.  Specifically, most holographic models assume bulk fields that satisfy $\m^2 =0$ either by choosing $\xi \sim - \eta^{-1}$ or taking the massless minimal coupled limit. For the latter, one can argue that quantum effects, which we regard as thermal, could spontaneously break the dS symmetry. For light-fields which satisfy $\m^2 \leq H^2$, the above approximation can be extended to the regime $a(\t_{\rm in}) \m < \tmk < H$. This corresponds to  modes that are initially subhorizon which exit the horizon, fueling as a result the thermal backreaction. 

Then, we can write the momentum integral as:
\bad
 {\rm Im} \left\{ \frac{1}{\beta}\int^\infty_0 d \tmk \  \frac{\beta \tmk}{e^{\b \tmk} -1} e^{-\tmk^2 A + i \tmk \Delta x } \right\}
\ead
and use the Bernoulli generating function:
\bad
\frac{t}{e^t -1} = \sum^\infty_{n =0} \frac{B_{n} t^{n}}{n!} 
\ead
with $B_{n}$ the even Bernoulli numbers, so that our integral is equal to the infinite sum:
\bad
{\rm Im} \left\{ \sum_{n =0}^\infty \frac{B_{n}}{n!} \b^{n-1} \int^\infty_0 d \tmk \ \tmk^{n} e^{-\tmk^2 A + i \tmk \Delta x } \right\} \ .
\ead
The odd Bernoulli numbers $B_{2n+1} = 0$, for $n \geq 1$, while $B_1 = - \frac{1}{2}$. 
The integral converges only when $\beta \tmk <2 \pi$ which corresponds to $\tmk < H$. This condition serves as a cutoff that physically 
corresponds at late-times to the superhorizon modes, which are the ones responsible for the primordial fluctuations anyways. As long as $H$ is large, 
the residual part of the integral that corresponds to subhorizon modes $\tmk \geq H$, vanishes due to the Gaussian exponential decay. 
Then, since $B_{n} (2 \pi)^{n} /(n)!  = (-1)^{1+\frac{n}{2}} \zeta(n)$, where $\zeta(n)$ the Riemann zeta function, for $\beta = \beta_{\rm dS}$ the sum
\bad
\frac{1}{2 \pi}\sum_{n =0}^\infty \frac{B_{ n} (2 \pi)^{n}}{(n)!} H^{1-n} \int^\infty_0 d \tmk \ \tmk^{n} e^{-\tmk^2 A + i \tmk \Delta x }  \ ,
\ead 
converges when $n \rightarrow \infty$. 
Furthermore, it is straightforward to show that the imaginary part of the $\tmk$-integral is equivalent to Kummer's confluent hypergeometric function ${_1F_1}[a,b,z]$ so that we obtain 
\bad
\frac{|\Delta \tx|}{4 \pi}\sum_{n =0}^\infty \frac{B_{n} (2 \pi)^{n}}  {n!} \Gamma\left(1 + \frac{n}{2} \right) H^{1-n}   
\ A^{-1-\frac{n}{2}} e^{-\frac{\Delta \tx^2}{4A}} {_1F_1}\left( \frac{1-n}{2} , \frac{3}{2}, \frac{\Delta x^2}{4 A}\right) \ .
\ead 
Then, by using the series representation \cite{Abramowitz}
\bad
{_1}F_1 [a,b,z]  = \sum^\infty_{m=0} \frac{(a)_m }{(b)_m} \frac{z^m}{m!} \,
\ead 
\eq{Iint} becomes:
\begin{align}
I(x_1, x_2) = \frac{|\Delta \tx |}{4 \pi} & \sum_{n =0}^\infty \frac{B_{n} (2 \pi)^{n}}  {n!}  \Gamma\left(1 + \frac{n}{2} \right) H^{1-n}  \sum^\infty_{m=0} \frac{2^{-2m}}{m!} \frac{(\frac{1-n}{2})_m}{(\frac{3}{2})_m } \Delta \tx^{2m} \times \nn \\
& \times \int^\infty_{-\infty} dT e^{-2 \nu T} \left[ \tau_1 \tau_2 \cosh 2T - \frac{1}{2} (\t^2_1 + \t^2_2)\right]^{-1 - \frac{n}{2} - m} \int^\infty_0 dv \ v^{\frac{n}{2}+m} e^{-B v}  
\end{align}
with 
\bad \label{Bparamet}
B(x_1,x_2) = \frac{1}{2} + \frac{\Delta x^2}{4 \left(|\t_1 \t_2| \cosh 2T - \frac{1}{2} (\t_1^2 + \t_2^2) \right)} {>0}
\ead
and $\Delta s^2 = (\t_1 - \t_2 )^2  - \Delta \tx^2$. For the last two integrals, one can follow a similar process as described in Appendix \ref{Point-splitting}, where the relation
\bad
\int^\infty_{-\infty} dT e^{-2 \nu T} &  \left[ \tau_1 \tau_2 \cosh 2T - \frac{1}{2} (\t^2_1 + \t^2_2)\right]^{-1 - \rho}  \int^\infty_0 dv \ v^{\rho} e^{-B v}   \\
&= \pi^{\frac{1}{2}} |\t_1 \t_2|^{-1-\rho} \frac{\Gamma[1+\rho+ \nu] \Gamma[1+\rho-\nu]}{ \Gamma[\frac{3}{2} + \rho ]}  F\left(1+\rho +\nu, 1 + \rho -\nu, \frac{3}{2} + \rho; \frac{\chi+1}{2}\right) \ ,
\ead
is used, so that for $\rho =\frac{n}{2}+  m $, we obtain
\bad \label{Iint2}
I(x_1, x_2) &= \frac{|\Delta \tx |}{4 \pi^{\frac{1}{2}}} \sum_{n =0}^\infty \frac{B_{n} (2 \pi)^{n}}  {n!} \Gamma\left(1 + 
\frac{n}{2} \right) H^{1-n}  \sum^\infty_{m=0} \frac{2^{-2m}}{m!} \frac{(\frac{1-n}{2})_m}{(\frac{3}{2})_m } |\t_1 \t_2|^{-1 -\rho} \Delta \tx^{2m} \times \\
&  \qquad \times \frac{\Gamma[1+\rho+ \nu] \Gamma[1+\rho-\nu]}{ \Gamma[\frac{3}{2} + \rho ]}  F\left(1+\rho +\nu, 1 + \rho -\nu, \frac{3}{2} + \rho; \frac{\chi+1}{2}\right)\, .
\ead 
Then from \eq{dG2} we get
\bad \label{dG3}
\delta G(x_1 - x_2) &=  \frac{H^3}{8 \pi^{7/2} } |\t_1 \t_2 |^{1/2}  \sum_{n =0}^\infty \frac{B_{n} (2 \pi)^{n}}  {n!} 
\Gamma\left(1 + \frac{n}{2} \right) H^{-n}  \sum^\infty_{m=0} \frac{2^{-2m}}{m!} \frac{(\frac{1-n}{2})_m}{(\frac{3}{2})_m } |\t_1 \t_2|^{-\rho} \Delta \tx^{2m} \times \\
&  \qquad \times \frac{\Gamma[1+\rho+ \nu] \Gamma[1+\rho-\nu]}{ \Gamma[\frac{3}{2} + \rho ]}  F\left(1+\rho +\nu, 1 + \rho -\nu, \frac{3}{2} + \rho; \frac{\chi+1}{2}\right)  \ .
\ead
As we have already argued, the result ought to converge away from $\chi = 1$, even though it 
contains two infinite series. In particular, we can think of the $n$-sum as an expansion with respect to the dimensionless parameter 
\bad
H^{1-n} |\t_1 \t_2|^{\frac{1-n}{2}}  =  \left(H^2 \t_1 \t_2 \right)^{\frac{1-n}{2}} \ ,
\ead
and since we have assumed that $H$ is large and $\t_1,\t_2$ arbitrary time-instances, this enables us to truncate to order $n= 1$. Then, we obtain in leading order $n=0$:
\bad \label{dG4}
\delta G(x_1 - x_2) &=  \frac{H^3}{8 \pi^{7/2}} |\t_1 \t_2 |^{1/2}   \sum^\infty_{m=0} \frac{2^{-2m}}{m!} \frac{(\frac{1}{2})_m}{(\frac{3}{2})_m }  \left( \frac{\Delta \tx^{2}}{|\t_1 \t_2|} \right)^m \times \\
&  \qquad \times \frac{\Gamma[1+m+ \nu] \Gamma[1+m-\nu]}{ \Gamma[\frac{3}{2} + m ]}  F\left(1+m +\nu, 1 + m -\nu; \frac{3}{2} + m; \frac{\chi+1}{2}\right) + {\cal O} (H^{2} ) 
\ead
with the next to leading order (NLO) term ($n=1$) for $m=0$:
\bad
\delta G^{\rm NLO}(x_1 - x_2) &=  -\frac{H^2}{16 \pi^{2} } \Gamma\left(\frac{3}{2}+ \nu\right) \Gamma\left(\frac{3}{2}-\nu\right) F\left(\frac{3}{2}+ \nu, \frac{3}{2}- \nu; 2 ; \frac{\chi+1}{2}\right)\, ,
\ead
cancelling exactly the zero-temperature BD result \eq{GBD}. This suggests that the thermal state contains the final information of the system after the tracing out performed by the TFD rotation. Notice that at the coincident point limit $\Delta \tx \rightarrow 0$, the residual $n=1, m \neq 0$ terms vanish. Thus, having the calculation of the EMT in mind, we choose to discard them. Then, the full thermal Wightman function is given by 
\bad \label{Gb}
G_\b(x_1 - x_2) &=  \frac{H^3}{8 \pi^{7/2}} |\t_1 \t_2 |^{1/2}   \sum^\infty_{m=0} \frac{2^{-2m}}{m!} \frac{(\frac{1}{2})_m}{(\frac{3}{2})_m }  \left( \frac{\Delta \tx^{2}}{|\t_1 \t_2|} \right)^m \times \\
&  \qquad \times \frac{\Gamma[1+m+ \nu] \Gamma[1+m-\nu]}{ \Gamma[\frac{3}{2} + m ]}  F\left(1+m +\nu, 1 + m -\nu; \frac{3}{2} + m; \frac{\chi+1}{2}\right) + {\cal O} (H) 
\ead
which naturally satisfies the equation of motion:
\bad
\Box_1 G_\b(x_1 - x_2) - \mu^2 G_\b(x_1 - x_2)  = \Box_2 \delta G_\b(x_1 - x_2) - \mu^2\delta G_\b(x_1 - x_2) = 0,
\ead 
where the index on the d'Alambertian denotes the coordinate being differentiated. The way to prove that 
$G_\b$ satisfies the above condition is to write the left-hand side as a polynomial of $\Delta \tx^2$ and use the hypergeometric differential equation
\bad \label{HDE}
z(1-z) \frac{d^2}{dz^2} F(a,b;c;z) + \left(c-(a+b+1)z\right) \frac{d}{dz} F(a,b,;c;z) - a b F(a,;c;z) = 0
\ead
individually for every order. Furthermore, it is possible to foresee that since soon we will be forced to 
take the limit $|\Delta \tx| \rightarrow 0$ in order to obtain the renormalized EMT, the sufficient consistency check for a conserved EMT is at $|\Delta \tx|^0$.  
Since \eq{Gb} turned out to have the form of a hypergeometric function yet with its arguments shifted by $-\frac{1}{2} +m$, 
we can calculate its covariant derivatives by following the same pattern as in Appendix \ref{Point-splitting}. Nevertheless, we need 
to be extra cautious, due to the remnant $|\t_1 \t_2 |^{1/2 -m}$ term, which will complicate the time-derivatives. 

As BD argued, it does not make any difference if we first renormalize $G_\b$ and then compute the thermal EMT or 
if we go the other way around. Here, we will follow the latter steps, keeping in mind that $F(a,b,c;z)$ diverges when $\chi \rightarrow 1$. 
In particular, near the specific singular point for $a = 1 + m + \n,\, b = 1 + m -\n$ and $c = \frac{3}{2} + m $, the solution to the hypergeometric 
differential equation \eq{HDE} can be written as the linear transformation \cite{Abramowitz}:
\bad \label{Hypergeom_near_1}
F  \left(1+m+\n, 1+ m- \n, \frac{3}{2} + m,\frac{\chi+1}{2}\right) &= \\
  \frac{\Gamma \left(\frac{3}{2} + m \right) \Gamma \left( -\frac{1}{2} - m \right)}{\Gamma \left( \frac{1}{2} + \nu \right) 
  \Gamma \left( \frac{1}{2} - \n \right)} &  F \left(1+m+\n, 1+m- \n, \frac{3}{2} + m, \frac{1-\chi}{2}\right) \\
 \qquad + \left( \frac{1-\chi}{2} \right)^{-1/2 -m}  &\frac{\Gamma \left(m+ \frac{3}{2}\right) \Gamma \left(m+ \frac{1}{2}\right) }
 {\Gamma \left( 1+ m + \n \right) \Gamma \left(1+ m - \n \right) }  \times  \\
&  \qquad \times  F \left(\frac{1}{2}-\n, \frac{1}{2}+ \n, \frac{1}{2} - m, \frac{1-\chi}{2}\right)
\ead
with the divergences located in the second term. In order to regulate these, the point-splitting technique allows us to express 
$1-\chi$ in terms of the proper-distance $\epsilon$ of the two spacetime points $x_1 (-\epsilon),x_2(\epsilon)$. 
Then, we can think of the double limit $x_{1,2} \rightarrow x$ as a parallel transportation fig. \ref{Fig_Parallel_Transport} of 
each distinct point towards the middle, assuming that both belong to the same geodesic curve. Then from 
\eq{expz}, $1-\chi \sim {\cal O} (\epsilon^2)$ which means that in \eq{Hypergeom_near_1}, the first term contains only even 
powers of $\epsilon$ while the second contains negative and odd powers.
As in \cite{bunch1978}, we subtract the divergent terms from the bare result while 
making sure that the final EMT remains conserved. These divergent pieces can then be used to renormalize bare constants 
such as Newton's constant $G$ via the introduction of counter-terms in the gravitational action, as the Schwinger-DeWitt process suggests. 
Subsequently, by subtracting the negative powers of $1-\chi$ from \eq{Hypergeom_near_1}, the  non-zero terms remaining in \eq{Gb} at the $\epsilon \rightarrow 0$ limit come from the expansion:
\bad
F \left(1+m+\n, 1+m- \n, \frac{3}{2} + m, \frac{1-\chi}{2}\right) &= \frac{\Gamma \left(\frac{3}{2} + m\right)}{\Gamma \left(1 + m + \n\right) \Gamma \left(1+  - \n \right)} \times \\
& \qquad \times \sum^\infty_{l = 0} \frac{\Gamma \left(1 + m + \n + l \right) \Gamma \left( 1 + m -\n + l \right)}{\Gamma \left( \frac{3}{2} + m + l \right)} \frac{(1-\chi)^l}{2^l l!}
\ead   
so that
\bad \label{Hypergeom_Limit}
 \lim_{\epsilon \rightarrow 0} \left[ F \left(1+m+\n, 1+m- \n, \frac{3}{2} + m, \frac{\chi+1}{2}\right) - {\cal O} (\epsilon^{-1} ) \right] = \frac{\Gamma \left(\frac{3}{2} + m \right) \Gamma \left( -\frac{1}{2} - m \right)}{\Gamma \left( \frac{1}{2} + \nu \right) \Gamma \left( \frac{1}{2} -\nu \right) } \ .
\ead 
Moving on, we will proceed to calculate the thermal EMT as the BD case of \eq{Tmnb2}:
\bad \label{VEVTb}
\braket{T^{\rm BD}_{\mu \nu}(x)}_{\b} &=  \lim_{\epsilon \rightarrow 0} \frac{1}{2} (1 - 2 \xi) \Bigl(\nabla_{\mu,1} \nabla_{\nu,2} G_{\b}  
+ \nabla_{\mu,2} \nabla_{\nu,1}  G_{\b} \Bigr) + \left(- \frac{1}{2} + 2 \xi \right) g_{\m \n} \nabla_{\kappa,1} \nabla^{\kappa,2} G_{\b}  \\
& - \xi  \Bigl(\nabla_{\mu,1} \nabla_{\nu,1} G_{\b} + \nabla_{\mu,2} \nabla_{\nu,2} G_{\b} \Bigr)  + 
\xi g_{\mu \nu}  \Bigl(\nabla_{\kappa,1} \nabla^{\kappa,1} G_{\b} +\nabla_{\kappa,2} \nabla^{\kappa,2} G_{\b} \Bigr)  \\
& \qquad -\frac{1}{2} m^2 g_{\mu \nu} G_{\rm BD} + \xi G_{\mu \nu} G_{\b}\, ,
\ead   
using the chain rule in order to calculate the various derivatives of the  thermal Wightman function \eq{Gb}. In the relevant limit, the non-zero derivatives of $z\equiv 1+\chi/2$ are
\begin{subequations}
\bad
\lim_{\epsilon \rightarrow 0} \frac{d^2 z}{d \t_1^2} = \lim_{\epsilon \rightarrow 0} \frac{d^2 z}{d \t_2^2} = -\lim_{\epsilon \rightarrow 0} \frac{d^2 z}{d \t_1 d \t_2} = \frac{1}{2 \t^2}
\ead  
and
\bad
\lim_{\epsilon \rightarrow 0} \frac{d^2 z}{d x^i_1 d x^j_1} = \lim_{\epsilon \rightarrow 0} \frac{d^2 z}{d x^i_2 d x^j_2} = -\lim_{\epsilon \rightarrow 0} \frac{d^2 z}{d x^i_1 d x^j_2} = -\frac{1}{2 \t^2} \delta_{ij}
\ead
\end{subequations}
with $i,j=1,2,3$ the spatial indices. Each double covariant derivative contains the non-zero 4d dS Christoffel symbols
\bad
\Gamma^0_{\mu \nu} = -\frac{1}{\t} \delta_{\mu \nu}, \qquad \Gamma^\mu_{0 \nu} = -\frac{1}{\t} \delta_\n^\m  \,
\ead
and by using the identity \eq{dFdz} along with \eq{Hypergeom_near_1} and \eq{Hypergeom_Limit}, we can straightforwardly compute each component of \eq{VEVTb} separately. 

The process is analogous to the BD EMT computed in Appendix \ref{Point-splitting} so here we will just present the results. 
As expected, the thermal EMT remain diagonal with the time and spatial components:
\begin{subequations} \label{VEVTb2}
\bad 
\braket{T^{\rm BD}_{00}(x)}_{\b} = -\frac{H^3 }{12 \pi^3} \frac{1}{|\t|} \frac{\Gamma(1+ \n) \Gamma(1-\n)}{\Gamma \left( \frac{1}{2}+ \nu\right) \Gamma \left( 
\frac{1}{2} - \nu\right)}  \left(1 + 12 \eta^{-1} - 6 \xi \right) 
\ead 
 \bad
\braket{T^{\rm BD}_{ij}(x)}_{\b} = \delta_{ij} \frac{H^3 }{18 \pi^3} \frac{1}{|\t|} \frac{\Gamma(1+ \n) \Gamma(1-\n)}{\Gamma \left( \frac{1}{2}+ \nu\right) \Gamma \left( 
\frac{1}{2} - \nu\right)}  \left(1 + 12 \eta^{-1} - 6 \xi \right) 
\ead 
\end{subequations}
which is evidently conserved. Here we have used again the dimensionless parameter $\eta$ which in our approximation should be large. 
Since we expect $\nu$ to obtain half integer values, e.g. the scale invariant invariant case 
corresponds to $\nu = 3/2$, the function $\Gamma\left(\frac{1}{2} - \n\right)$ in the denominator diverges as $(\nu - \frac{3}{2})^{-1}$. This results into the 
vanishing of both \eq{VEVTb2}, which means that no leading order thermal corrections are expected when $\nu = 3/2$. 
As a result, no thermal backreaction will be detected in this case. The same happens for the conformal coupling $\xi = \frac{1}{6}$ 
which corresponds to $\nu = \frac{1}{2}$. 
In both cases, we seem to reach the conclusion that by choosing $m^2=0$ and either $\xi=0$ or $\xi = \frac{1}{6}$, no thermal corrections are observed.\footnote{
From a slightly different point of view, one of the conclusions in \cite{FotisAntonis2} was also that the thermal deviation 
of the scalar spectral index $n_S$ tends to zero as the scale invariant limit is approached. }
Using Euler's reflection formula
\bad
\frac{\Gamma(1+ \n) \Gamma(1-\n)}{\Gamma \left( \frac{1}{2}+ \nu\right) \Gamma \left( \frac{1}{2} - \nu\right)} = \nu \cot \left( \pi \nu \right)\, ,
\ead
\eq{VEVTb2} can be further simplified as
\bea\label{VEVTb3}
\braket{T^{\rm BD}_{00}(x)}_{\b} &=& -\frac{H^3}{12 \pi^3} \frac{1}{|\t|} \nu \cot \left( \pi \nu \right)  \left(1 + 12 \eta^{-1} - 6 \xi \right) \\
\braket{T^{\rm BD}_{ij}(x)}_{\b} &=& \delta_{ij} \frac{H^3 }{18 \pi^3} \frac{1}{|\t|} \nu \cot \left( \pi \nu \right)  \left(1 + 12 \eta^{-1} - 6 \xi \right) \, .
\eea 
Finally, in contrast to \cite{bunch1978}, since our dimensionless parameter $\eta$ is assumed to be large, we can expand 
$\nu = \sqrt{\frac{d^2}{4} - \frac{\mu^2}{H^2}}$ in its inverse powers, up to $O(\eta^{-1})$ for simplicity. Then, \eq{VEVTb3} becomes
\bea  \label{VEVTb4}
\braket{T^{\rm BD}_{00}(x)}_{\b} &=& -\frac{H^3 }{12 \pi^3} \frac{1}{|\t|}  \left({\cal C}_1 - 6 {\cal C}_1 \xi + (12{\cal C}_1  - {\cal C}_2 + 6 \xi {\cal C}_2 )\eta^{-1}  \right) \\
\braket{T^{\rm BD}_{ij}(x)}_{\b} &=& \delta_{ij} \frac{H^3 }{18 \pi^3} \frac{1}{|\t|} \left({\cal C}_1 - 6 {\cal C}_1 \xi + (12{\cal C}_1  - {\cal C}_2 + 6 \xi {\cal C}_2 )\eta^{-1}  \right)  \, ,
\eea
where
\bea \label{C1}
{\cal C}_1 &=& \sqrt{\frac{9}{4} - 12 \xi} \cot \left(\pi \sqrt{\frac{9}{4} - 12 \xi}  \right)\\
{\cal C}_2 &=& - 6 \pi + \frac{4 \sqrt{3}}{\sqrt{3 - 16 \xi}} \cot \left(\pi \sqrt{\frac{9}{4} - 12 \xi}  \right) - 6 \pi \cot^2 \left(\pi \sqrt{\frac{9}{4} - 12 \xi}  \right)
\eea
are the 0th and 1st order coefficients in the expansion of $\nu  \cot(\pi \nu) = {\cal C}_1 - {\cal C}_2 \eta^{-1} + {\cal O} (\eta^{-2})$. 

\section{The thermal backreaction} \label{thermal backreaction}

We are now ready to compute the thermal backreaction which is the main goal of this paper. The proper way to do it is to solve the system of ODE's:
\begin{subequations}
\bad \label{ODE1}
\phi'' + 2 \frac{\tilde a'}{\tilde a} \phi' + \tilde a^2 \mu^2 \phi = 0
\ead
\bad \label{ODE2}
\tilde H^2 = H^2  + \frac{8 \pi G}{3} \frac{1}{\tilde a^2 } \braket{T_{00}}_{\b}
\ead
\bad \label{ODE3}
\frac{(\tilde a')^2}{\tilde a^4} - 2 \frac{\tilde a''}{\tilde a^3}  =- 3 H^2 +8 \pi G \frac{1}{\tilde a^2} \braket{T_{11}}_\b 
\ead
\end{subequations}
for the new metric $\tilde a (\t) = a(\t) + \delta a(\t)$ (correspondingly, we define $\tilde H = \frac{{\tilde a}'(\t)}{{\tilde a}^2 (\t)}$ as the new Hubble constant)
and the scalar field $\phi$ simultaneously, which is usually done numerically since 
only a few analytic solutions are known for specific cases. Here $G = 1/ M^2_{\rm pl}$ and we assume $\tilde a$ to be of the FRW form 
\bad
ds^2 = \tilde a^2(\t) \left(- d \t^2 + d \tx^2 \right)
\ead 
since the thermal EMT \eq{VEVTb4} is only time-dependent and diagonal. 
Nevertheless, perturbative solutions have been found \cite{Starobinsky1980} that suggest that the universe was initially in a dS state with constant curvature, 
which is also our starting point. In addition, a further analysis has been performed \cite{Anderson1986} for the case of conformally 
invariant free quantum fields either with or without classical radiation or conformally coupled fields. There, the effective Hubble constant 
was computed for various scenarios concerning the initial particle spectrum. 

Here we will instead apply the iteration method where we first solve the eom \eq{ODE1} for fixed dS background and then substitute 
the results into the two Friedmann equations \eq{ODE2} and \eq{ODE3}. Then, the resulting $\delta a$ should correspond to the 
first order in $G$ deviation away from the dS geometry, caused by the appearance of the scalar field. As a result, we define the energy $\delta \rho \equiv - \braket{T^0_0}$ and pressure $\delta p \equiv \frac{1}{3} T^i_i$ densities, which alter the initial vacuum energy $\rho >0$ and pressure $p <0$ that drive inflation.   

The solution to \eq{ODE2}, \eq{ODE3} once we substitute in \eq{VEVTb4} is 
\bad
 \tilde a(\t)  = a(\tau) \left[ 1 + \frac{1}{18 \pi^2} G H^3 |\t| \nu \cot \left( \pi \nu \right)  \left(1 + 12 \eta^{-1} - 6 \xi \right) + {\cal O} (G^2) \right]
\ead
which in turn introduces the 1st order correction
\bad \label{Hdeviation}
\tilde H = H \left[1 - \frac{1}{9 \pi^2} G H^3 |\t|  \nu \cot \left( \pi \nu \right)  \left(1 + 12 \eta^{-1} - 6 \xi \right) + {\cal O} (G^2)\right] \ 
\ead 
to the Hubble constant. Then, the deformed expansion rate $\tilde {\cal H} \equiv \tilde a'/\tilde a = \tilde a \tilde H $ is 
\bad \label{ExpRate}
\tilde {\cal H} = a H \left[ 1-  \frac{1}{18 \pi^2} G H^3 |\tau|  \nu \cot \left( \pi \nu \right)  \left(1 + 12 \eta^{-1} - 6 \xi \right) \right] .
\ead  
As long as $\xi < \frac{1}{6} + 2 \eta^{-1}$, the sign of the thermal corrections is fixed by $\cot (\pi \nu)$ . Near the massless limit $\eta^{-1} \rightarrow 0$ where $\cot (\pi \nu) \sim {\cal C}_1$ from \eqref{C1}, we will restrict our analysis with the double inequality
\bad
\frac{1}{2} \leq \nu \leq \frac{3}{2}
\ead
with the two extrema corresponding to a vanishing thermal effect. We interpret this as follows. For $\xi =0 \Rightarrow \nu = \frac{3}{2}$, the particle bath does not explicitly couple with gravity, and thus should not feel any curvature effects and vice versa. On the other hand, for $\xi = \frac{1}{6} \Rightarrow \nu = \frac{1}{2}$ conformal symmetry is restored, hence a comoving observer in conformally flat coordinates should not feel any particle production in the first place, since the latter would break scale invariance.  

When $\frac{1}{2} \leq \nu < 1$, ${\cal C}_1 < 0$, which results to an increase in the expansion rate \eqref{ExpRate}. Consequently,  there is a shift in the time of horizon exit which enables some additional modes to become superhorizon. Furthermore, the energy and pressure deviations become positive and negative respectively, indicating a further accelerated expansion. In contrast, for $1 < \nu \leq \frac{3}{2}$ we have exactly the opposite effects. Now ${\cal C}_1 >0$ so that the expansion rate decreases and  some modes which where previously superhorizon now re-enter the causal patch. The reinsertion of these extra modes, should further produce thermal fluctuations of first order in $G$ to the power spectrum, along with the ones showcased in \cite{FotisAntonis2}. The vacuum energy is reduced since $\delta \rho <0$ while $\delta p >0$, thus the thermal backreaction slows the expansion. 

As $\nu \rightarrow 1^\pm$, the above effects amplify in both cases and in order to describe a physically accepted system, a cut-off for $\nu$ should be introduced. A reasonable bound should be read from \eq{ExpRate}. Since we want to interpret the thermal effects as fluctuations to the expansion of the universe, they should neither behave like or cancel the effect of the cosmological constant. Hence, we suppose that in general
\bad
 GH^3 |\tau | | \nu \cot \left( \pi \nu \right)  \left(1 + 12 \eta^{-1} - 6 \xi \right)  | <1 
\ead  
which is a non trivial inequality, as it contains two variables in terms of $m$ and $\xi$. In the massless limit, the above inequality reduces to $G H^3 |\tau| |{\cal C}_{1} | < 1$, which has no known analytic solutions for $\xi$, while its numerical analysis lies beyond the scope of this work. Notice that when $\xi > \frac{5}{48}$, we obtain a tachyonic mass which we consider to be unphysical. In the opposite regime, as $\xi \rightarrow 0$ and further down, the mass parameter becomes comparable to the dS scale and most slow-roll models would produce an inadequate inflationary period. In any case, these strong backreaction effects cannot be described by the iteration method used in this current work and thus \eq{Hdeviation} should be regarded as an effective result away from the $\n \rightarrow 1$ limit. 

\begin{figure}[h] 
\centering
\includegraphics[scale=0.55]{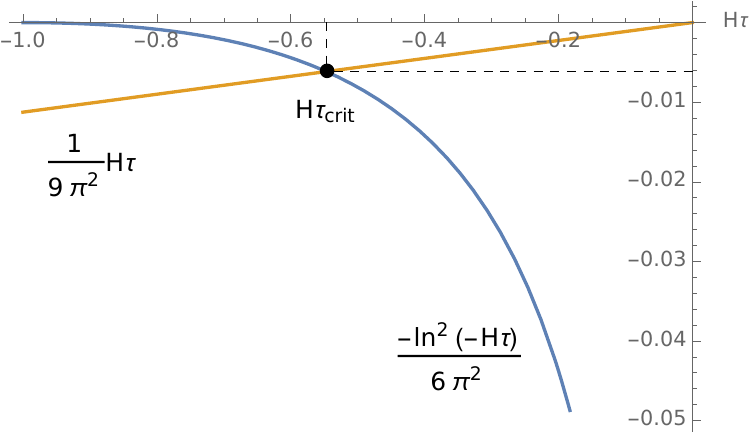}
\caption{The thermal and graviton loop correction for $\xi =0, m^2 = G H^4$. The (orange) line corresponds to the thermal effect while the (blue) curve represents the graviton correction.
The square of the logarithm of the latter starts to dominate at late times right after $H \tau_{\rm crit} \simeq  -0.547$.} \label{WoodardvsUs}
\end{figure}

The thermal corrections to $H$ can be contrasted with the the ones calculated in \cite{WoodardTsamis1996}, 
where the quantum backreaction was considered due to the 2-loop graviton VEV. It was concluded that those loop effects shift the Hubble constant to second order in $G$ as:
\bad \label{WoodardTsamis}
\tilde H  = H \left[1 -  \frac{1}{ \pi^2} G^2 H^4 \left(\frac{1}{6} \ln^2 (- H \t )  + {\rm subleading} \right) + {\cal O} (G^3) \right] .
\ead
At first sight, while \eq{Hdeviation} seems to contain a significantly larger shift due to it being of lower order, such a conclusion still depends on the 
mass of the scalar field and the conformal coupling $\xi$. For instance, as mentioned already, in either the minimally or conformally coupled cases 
$\xi = 0, \frac{1}{6}$ the effects vanish completely at the massless limit. For large $\eta$, this is still the case for $\xi = \frac{1}{6}$ whereas for $\xi = 0$:  
\bad
\nu \cot \left( \pi \nu \right)  \left(1 + 12 \eta^{-1}\right) = 6 \pi  \eta^{-1} + {\cal O} (\eta^{-2} )  
\ead
which could drop our calculation one order depending on whether $m^2 > G H^4$ or not. Moreover, the time $\tau$ in which 
the backreaction is observed plays also a significant role when comparing the two corrections. Specifically, at the dS time 
scale $H |\tau| \rightarrow 1$ the logarithmic correction in \eq{WoodardTsamis} vanishes though it starts to dominate the 
closer we get to $\tau \rightarrow 0$ where our result is negligible. In order to further illustrate the two corrections compared to each other, 
let us assume $\xi =0$ and $m^2 = GH^4$. Then, at the critical time instance  $H \tau_{\rm crit} \simeq  -0.547$ and afterwards, 
the square of the logarithm begins to dominate as illustrated in \fig{WoodardvsUs}. 

Gravitational fluctuations in the Einstein equation of the type discussed here have been analyzed before. 
For example recently in \cite{Bradenberger2023} (also references therein), the second order perturbations were shown to impose a backreaction on the expansion rate:
\bad\label{Brand}
\frac{\delta {\cal H}}{\cal H} \sim  - (1+ w) \frac{\delta \rho^2}{\rho^2}\, ,
\ead
with $\rho, \delta \rho$ the spatially averaged energy density and its variation, 
as observed by a certain local observer. This provides further evidence that the choice of initial conditions affects the sign of the change in ${\cal H}$. In the case of \eq{BDVEV}, due to the dS isometry, the equation of state is $w= - 1$ (see \eq{BDTdSiso})
and the above gravitational fluctuations vanish, as they should. This means that the quantum BD backreaction (the last term in \eq{BDbackreact}), 
is purely matter-like and that is why it is positive. 
Our thermal effects instead, which from \eq{VEVTb4} can be seen to correspond to $w = - \frac{2}{3}$, for ${\cal C}_1 > 0 $ can be seen to also have a gravity-like backreaction interpretation.
The total effect, as interpreted from the point of view of \eq{Brand}, seems to be an effective sum of vacuum energy and pure radiation, with
$\frac{\delta \rho^2}{\rho^2} \sim \frac{2}{45\pi^2} G H^3 |\tau| \nu \cot \left( \pi \nu \right)  \left(1 + 12 \eta^{-1} - 6 \xi \right)$.

A few final comments.
If we assume that $\phi$ has a classical part, then there will be also a classical backreaction which is calculated explicitly in Appendix \ref{Zero-temp Backreaction}. Then, 
the combined classical plus thermal effect is simply the sum of the two effects:
\bad \label{HdeviationClassThermal}
\tilde H = H \left[1+ 8 \pi  G \phi^2_0  \tau^{3 + 2 \nu}  \frac{A(\eta)}{\eta}  -  \frac{1}{9 \pi^2} G H^3 |\t|  \nu \cot \left( \pi \nu \right)  \left(1 + 12 \eta^{-1} - 6 \xi \right) + {\cal O} (G^2)\right] 
\ead 
with $A(\eta)$ given in \eq{A(eta)}. If we regard $\phi$ as the inflaton, the thermal generalization of the slow-roll parameter $\epsilon$ which controls how the spacetime expansion recedes, is:
\bad
\tilde \epsilon = \epsilon +\frac{\tilde H'}{\tilde a \tilde H^2} = \epsilon - \frac{1}{9 \pi^2} G H^3  |\t|  \nu \cot \left( \pi \nu \right)  \left(1 + 12 \eta^{-1} - 6 \xi \right)   \ .
\ead 
Seemingly, $0<\tilde \epsilon < 1$ meaning that while the thermal corrections are not capable of ending inflation, at least they do not break the basic slow roll assumption.  
From the mode function point of view, the new scale factor $\tilde a$ can be re-inserted into the eom \eq{ODE1} which produces the extra friction and mass terms:
\begin{subequations}
\bad
2 \frac{\tilde a'}{\tilde a} u'_\tmk = 2 \frac{ a'}{a}\left(1  - \frac{1}{18 \pi^2} G H^3 |\t|  \nu \cot \left( \pi \nu \right)  \left(1 + 12 \eta^{-1} - 6 \xi \right) + {\cal O}(G^2)  \right) u'_\tmk     
\ead
\bad
\tilde a^2 \mu^2 u_\tmk = a^2 \mu^2 \left(1 + \frac{1}{9 \pi^2} G H^3 |\t|  \nu \cot \left( \pi \nu \right)  \left(1 + 12 \eta^{-1} - 6 \xi \right)     + {\cal O}(G^2) \right) u_\tmk  \ .
\ead
\end{subequations}
Then, depending on whether $\nu \cot \left( \pi \nu \right)  \left(1 + 12 \eta^{-1} - 6 \xi \right) >0 $ or not, we can observe either a decrease in friction accompanied by an increase in the potential, or vice versa. 
These effects can be interpreted as slight reduction (or growth) in the damping of the oscillation of $\phi$, where vacuum (potential) energy is effectively traded for potential (vacuum) energy due to interactions with the thermal environment.
Let us note that one of the reasons we considered the simplest possible model i.e. one that contains only a quadratic potential 
was because it simplifies the implementation of the TFD formalism in curved spacetime. 
Of course, generalizations with interactions and/or more sophisticated field contents should be possible along the same lines. 
Finally, based on the above results, let us facilitate the distinction between curvature and thermal effects by stating that they are interchangeable and their interpretation depends on the observer's time evolution (static or comoving).

\section{Conclusions} \label{Conclusions}

In this work we have demonstrated how the TFD formalism can provide a convenient framework to study the intrinsic mechanism of particle production 
and the way it affects the propagation of a scalar field during the inflation epoch. In particular, we took advantage of the fact that a timelike observer ought to 
perceive the system being at a thermal equilibrium described by the dS temperature $T_{\rm dS} = H/2 \pi$. With that in mind, we considered a free, massive
scalar field located initially in the BD vacuum which we rotated into its thermal analog containing a distribution equivalent to the Bose-Einstein number density $n_{\rm B}$. 

Accordingly, we computed the 2-point function defined via the thermal state which could be decomposed into the original BD result plus a thermal correction, 
essentially expanding the work performed in \cite{bunch1978}. Interestingly, by assuming a sufficiently large $H$, the superhorizon modes where shown 
to contribute primarily to the latter, while the subhorizon part decays exponentially. The thermal corrections ultimately took the form of a double-sum, with the first sum 
given by $(H^2 \tau_1 \tau_2)^{1-n/2}$, which we effectively truncated at leading order $n=0$. 
An interesting aspect of the computation was that the next to leading order term $n=1$ 
was shown to cancel the BD result for $\Delta \tx \rightarrow 0$, so that the thermal Wightman function absorbed the BD contribution. 
The point-splitting technique was then used in order to locate and regulate the UV divergences which allowed us to compute the renormalized thermal EMT 
directly from the 2-point function in the coincident point limit. The result was expressed in terms of the (assumed to be) large dimensionless parameter 
$\eta = {\cal R}/m^2$ and shown to satisfy the conservation condition. In addition, it vanished at the non-minimal and conformal coupled massless cases, 
indicating that the corresponding state remains dS invariant.   
   
We solved the Einstein equations semiclassically for the thermal case and perturbatively found  a deformed metric that retains its 
FRW form, with its physical interpretation depending on the initial conditions. In particular, the sign of the corrections to the original dS scale factor and Hubble constant is determined by the mass of the scalar field and its explicit coupling to gravity. These corrections, either increase or decrease the Horizon size, causing some modes  to re-enter or exit the causal patch. To ensure that they behave like thermal fluctuations to the power spectrum, we imposed an upper bound on their magnitude. Following, we compared our work with the 2-loop graviton 
calculation of \cite{WoodardTsamis1996}. Although at first glance our corrections are of first order in $GH^2$ while the latter is of second order,  
we saw that the two can be, in principle, of the same magnitude. This depends on the mass of the scalar field, the value of the coupling $\xi$ and  
the time instance at which the calculations are performed. The closer the latter is to $\tau \rightarrow 0$, the graviton corrections increase 
while the thermal corrections become negligible. On the other hand, near the dS time scale $H |\t| \rightarrow 1$ it's the other way around with the thermal effects dominating. 
The deformed scale factor gives rise to extra terms in the equations of motion that modify the friction and mass terms in opposite directions. 
Physically, this will be translated to an earlier or later oscillatory period of $\phi$. By considering a classical field component as well, we computed the 
non-zero temperature slow-roll parameter $\tilde \epsilon$ which seems to still satisfy $\tilde \epsilon <1$, meaning that inflation persists in the presence of the thermal effects.

\section*{Acknowledgments}
The work of FK is supported in part by the National Natural Science
Foundation of China under grant No. 12342502.

\appendix

\section{The renormalized Bunch-Davies Energy-Momentum Tensor} \label{Point-splitting} 

In this section, we repeat the computation of the renormalized, zero-temperature or Bunch-Davies VEV of the Energy-Momentum tensor of \cite{bunch1978},
using the point splitting regularization technique.
The expression that yields the EMT is

\begin{align} \label{VEVTBD}
\braket{T^{\rm BD}_{\mu \nu}(x)}_{\rm BD} &=  \lim_{\substack{x_1 \rightarrow x \\ x_2  \rightarrow x}} \frac{1}{2} (1 - 2 \xi) \Bigl(\nabla_{\mu,1} \nabla_{\nu,2} G_{\rm BD}  + \nabla_{\mu,2} \nabla_{\nu,1}  G_{\rm BD} \Bigr) + \left(- \frac{1}{2} + 2 \xi \right) g_{\m \n} \nabla_{\kappa,1} \nabla^{\kappa,2} G_{\rm BD}  \nn \\
& - \xi  \Bigl(\nabla_{\mu,1} \nabla_{\nu,1} G_{\rm BD} + \nabla_{\mu,2} \nabla_{\nu,2} G_{\rm BD} \Bigr)  + \xi g_{\mu \nu}  \Bigl(\nabla_{\kappa,1} \nabla^{\kappa,1} G_{\rm BD} +\nabla_{\kappa,2} \nabla^{\kappa,2} G_{\rm BD} \Bigr)  \nn \\
& \qquad -\frac{1}{2} m^2 g_{\mu \nu} G_{\rm BD} + \xi G_{\mu \nu} G_{\rm BD}\, ,
\end{align}
where both fields and vacuum in $G_{\rm BD}$ are defined in the diagonal, BD basis as in \eq{GBDdiag}. As a slight deviation from \cite{bunch1978}, 
we will calculate the above expression directly via the derivatives of the BD Wightman function \eq{GBD} and then use the expansion of the 
hypergeometric function in order to regulate the result by subtracting the divergences. 
This differs from the original calculation where the expansion of the hypergeometric functions around $\epsilon \rightarrow 0$ was done first, 
so that the spacetime derivatives acted on the expanded terms. Basically, our way ensures generality keeping in mind that the 
thermal EMT originates from integrals that result in $F(a,b;c;z)$ with different arguments.

As a first step, we would like to arrive at \eq{GBD} by substituting \eq{ukBD} into \eq{GBD_0}. Doing so we obtain
\bea \label{GBD_1}
G_{\rm BD} (x_1 - x_2) &=& \frac{H^{d-1}}{2^{d+2} \pi^{d-1}} |\t_1 \t_2|^{d/2} g(\tau_1,\tau_2) \nonumber\\
g(\tau_1,\tau_2) &=& \int d^d \tk \ H^{(1)}_\nu \left(\tmk |\t_1| \right) H^{(2)}_\nu \left(\tmk |\t_2| \right) e^{i \tk \bm{\Delta x}}\, .
\eea
From now on we will set $d=3$ in order to specify our calculation to four dimensions. 
The double Hankel function integral is calculated by first charting the 3d momentum space in spherical coordinates and then by expressing 
their product as a double integral, through the identity
\bea \label{ProdHankelRep}
 H^{(1)}_\nu \left(\tmk |\t_1| \right) H^{(2)}_\nu \left(\tmk |\t_2| \right) &=& 
 \frac{2}{\pi^2} \int^{+\infty}_{- \infty} dT \  \int^{\infty}_0 \frac{dv}{v} \, I(v,T) \nonumber\\
 I(v,T) &=& \exp \left[ - 2 \nu T - \frac{\tmk^2 |\t_1| |\t_2| \cosh 2T}{v} - \frac{v}{2} + \frac{\tmk^2 |\t_2|^2 + \tmk^2 |\t_1|^2}{2v} \right]\, .\nonumber\\
\eea
Then, the $\tmk$ integral can be massaged into a Gaussian form so that
\be
g(\tau_1,\tau_2) = \frac{2}{\sqrt{\pi}}  \int^{+\infty}_{- \infty} dT  \left(|\t_1 \t_2| \cosh 2T + \frac{1}{2} (\t_1^2 + \t_2^2) \right)^{-3/2} \int^{\infty}_0 dv \  v^{1/2} e^{- vB}\, ,
\ee
with $B$ the same positive parameter given in \eq{Bparamet}. 
The $v$ integral is also Gaussian and it can be performed, to give
\bad
g(\tau_1,\tau_2) =
2^{3/2} |\t_1 \t_2|^{-3/2}  \int^{+\infty}_{0} dT \ \frac{\cosh \n T}{\left( \cosh2T - \chi \right)^{-3/2}}\, .
\ead
The last integral over $T$ can be written as an associated Legendre polynomial \cite{Erdelyi1}. Then, \eq{GBD_1} becomes
\bad
G_{\rm BD} (x_1 - x_2) = \frac{H^2}{8 \pi^2} \Gamma \left(\frac{3}{2} + \nu \right) \Gamma \left(\frac{3}{2} - \nu \right) (\chi^2 -1 )^{-1/2} P^{-1}_{\nu - \frac{1}{2}} ( -\chi).
\ead
This can be written in a more convenient form, in terms of a hypergeometric function using the identity \cite{Abramowitz}
\bad
(\chi^2 -1 )^{-1/2} P^{-1}_{\nu - \frac{1}{2}} ( -\chi) = \frac{1}{2} F \left( \frac{3}{2} + \n, \frac{3}{2} - \nu; 2 ; \frac{\chi+1}{2} \right)\, .
\ead
Here $\chi = 1     +  \frac{\Delta s^2}{2 \tau_1 \tau_2}$, where $\Delta s^2 = (\t_1 - \t_2)^2 - \Delta \tx^2$.
This gives the sought result \eq{GBD}: 
\bad \label{GBD_2}
G_{\rm BD} (x_1 - x_2) = \frac{H^2}{16 \pi^2}  \Gamma \left(\frac{3}{2} + \nu\right) \Gamma \left(\frac{3}{2}- \nu\right)  F \left( \frac{3}{2} + \nu,\frac{3}{2} - \nu;2;\frac{1}{2} (\chi + 1)\right). 
\ead
Notice that, when $x_1, x_2 \rightarrow x$, then $\chi \rightarrow 1$ and the hypergeometric function diverges. 
These are the hidden divergences that we have to take care of with the renormalization scheme. 

Now that we have succeeded in expressing $G_{\rm BD}$ as a hypergeometric function, we can proceed to compute its derivatives with respect 
to the spacetime points $x_1, x_2$. In particular, using the formula
\bad \label{dFdz}
\frac{d}{dz} F(a,b;c;z) = \frac{a \cdot b}{c} F(a+1,b+1;c+1;z)
\ead 
we obtain for the 00-component
\begin{subequations}
\be \label{GBD00}
\nabla_{0,1} \nabla_{0,2} \ G_{\rm BD}   = \frac{H^2}{64 \pi^2}  \Gamma \left(\frac{3}{2} + \nu\right) 
\Gamma \left(\frac{3}{2}- \nu\right) \left( \frac{9}{4} -\n^2 \right) \left\{ G^{(1)}_{\rm BD;00} + G^{(2)}_{\rm BD;00} \right\}\, ,
\ee
where
\bad
G^{(1)}_{\rm BD;00} = - \frac{1}{2 |\t_1 \t_2|}  F \left( \frac{5}{2} + \nu,\frac{5}{2} - \nu;3;\frac{1}{2} (\chi + 1)\right) 
\ead
and
\bad \label{G2BD00}
G^{(2)}_{\rm BD;00} &= - \frac{(|\t_1| - |\t_2|)^2 + (\tx_1 -\tx_2)^2}{4 \t^2_1 \t^2_2}  F \left( \frac{5}{2} + \nu,\frac{5}{2} - \nu;3;\frac{1}{2} (\chi + 1)\right) \\
& \qquad + \frac{1}{3} \left(\frac{25}{4} - \n^2 \right) \frac{\Delta \tx^4 - (\t^2_1 - \t^2_2)^2}{16 |\t_1|^3 |\t_2|^3}  F \left( \frac{7}{2} + \nu,\frac{7}{2} - \nu,4;\frac{1}{2} (\chi + 1)\right)\, .
\ead 
\end{subequations}
The contribution $G^{(1)}_{\rm BD;00}$ does not vanish in the limit $x_1,x_2 \rightarrow x$, while $\lim_{x_{1,2}\to x}G^{(2)}_{\rm BD;00} = 0$. Similarly, for the $ii$-component we obtain
\begin{subequations}
\bad \label{GBDii}
\nabla_{i,1} \nabla_{i,2} \ G_{\rm BD} = \frac{H^2}{64 \pi^2}  \Gamma \left(\frac{3}{2} + \nu\right) \Gamma \left(\frac{3}{2}- \nu\right) 
\left( \frac{9}{4} -\n^2 \right) \left\{ G^{(1)}_{{\rm BD};ii} + G^{(2)}_{{\rm BD};ii} \right\}
\ead
with
\bad
G^{(1)}_{{\rm BD};ii} =  \frac{1}{2 |\t_1 \t_2|}  F \left( \frac{5}{2} + \nu,\frac{5}{2} - \nu;3;\frac{1}{2} (\chi + 1)\right) = - G^{(1)}_{\rm BD;00} 
\ead
and
\bad \label{G2BDii}
G^{(2)}_{{\rm BD};ii} &= -  \frac{1}{3} \left(\frac{25}{4} - \n^2 \right) \frac{\Delta \tx^2}{4 |\t_1|^2 |\t_2|^2}  F \left( \frac{7}{2} + \nu,\frac{7}{2} - \nu;4;\frac{1}{2} (\chi + 1)\right)\, .
\ead 
\end{subequations}
Again, $\lim_{x_{1,2}\to x} G^{(2)}_{{\rm BD};ii} = 0$. Notice that
\bad \label{Double Derivative G}
\lim_{\substack{x_1 \rightarrow x \\ x_2  \rightarrow x}} \left[\nabla_{0,1} \nabla_{0,2} \ G_{\rm BD} + \nabla_{i,1} \nabla_{i,2} \ G_{\rm BD}\right] = 0
\ead
as the only non-zero terms are $G^{(1)}_{\rm BD;00}$ and $G^{(1)}_{{\rm BD};ii}$ which are exact opposites. Note that for the above expressions, the double index $ii$ refers to any spatial component given by $i=1,2,3$ and not to the sum of them. Moreover, 
\bad \label{Double Derivative G_2}
\lim_{\substack{x_1 \rightarrow x \\ x_2  \rightarrow x}} \nabla_{i,1} \nabla_{i,1} \ G_{\rm BD} = \lim_{\substack{x_1 \rightarrow x \\ 
x_2  \rightarrow x}} \nabla_{i,2} \nabla_{i,2} \ G_{\rm BD} = -\frac{H^2}{64 \pi^2}  \Gamma \left(\frac{3}{2} + \nu\right) 
\Gamma \left(\frac{3}{2}- \nu\right) \left( \frac{9}{4} -\n^2 \right)  \lim_{\substack{x_1 \rightarrow x \\ x_2  \rightarrow x}} G^{(1)}_{{\rm BD};ii}  
\ead  
a relation that will prove useful later on.  

\begin{figure}[h] 
\centering
\includegraphics[scale=0.8]{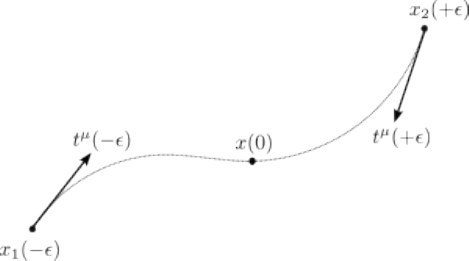}
\caption{The parallel transportation from the points $x_1$ and $x_2$ towards their middle point $x$ using the geodesic tangent vector $t^\mu$.} \label{Fig_Parallel_Transport}
\end{figure}

Now, we would like to show why $\lim_{x_{1,2}\to x} \{G^{(2)}_{\rm BD;00}$,\, $G^{(2)}_{{\rm BD};ii} \} = 0$. 
For this purpose, we first need to reparametrize the above expressions in terms of the proper-distance $\epsilon$ which separates 
$x_1, x_2$, see \fig{Fig_Parallel_Transport}, and then expand the hypergeometric functions around $\epsilon \rightarrow 0$. 
In general, the hypergeometric function $F(a,b,c;z)$ diverges when $z \rightarrow 1$. In our case
\bad \label{z of F}
z = \frac{\chi + 1}{2} \Rightarrow z - 1 = \frac{\Delta s^2}{4 \t_1  \t_2}\, ,
\ead
which means that we have to write $\frac{\Delta s^2}{4 \t_1  \t_2}$ as a power series in $\epsilon$ and throw away the negative powers. 
 
Assuming that $x_1, x_2$ belong to the same geodesic whose affine parameter is the proper distance $\epsilon$ with 
the points expressed in terms of the parameter as $x_1(-\epsilon)$ and $x_2(\epsilon)$, 
we want to parallel transport both towards the point $x (0)$. For simplicity, we will also assume that $x^\m_1 = x^\m_2$ for $\m = 2,3$ so that the 
only interesting directions will be those of the plane $\m=0,1$. Then we can Taylor expand around $x(0)$:
\begin{subequations}
\bad
\t_{1,2} = \t \mp \epsilon t^0_1 + \frac{1}{2}\epsilon^2 t^0_2 \mp \frac{1}{6} \epsilon^3 t^0_3 + {\cal O} (\epsilon^4) 
\ead
\bad
x_{1,2} = x \mp \epsilon t^1_1 + \frac{1}{2}\epsilon^2 t^1_2 \mp \frac{1}{6} \epsilon^3 t^1_3 + {\cal O} (\epsilon^4) 
\ead
\end{subequations}
where $t^\mu_1, t^\mu_2, t^\m_3, \cdots$ the expansion coefficients of the tangent vector $t^\m$. 
Using the dS geodesic equations, \cite{bunch1977} showed that $t^\mu_2, t^\m_3$ can be expressed in terms of $t^\mu_1$
and that the latter satisfy the constraint
\bad
(t^0_1)^2 - (t^1_1)^2 = H^2 \t^2 \Sigma \, ,
\ead
with $\Sigma = \pm 1$ depending on whether the geodesic is respectively space-like or time-like. Expanding the right-hand side of \eq{z of F} and its inverse gives:
\begin{subequations} \label{expz}
\bad
\frac{\Delta s^2}{4 \t_1  \t_2} = H^2 \epsilon^2 \Sigma + {\cal O}(\epsilon^4)
\ead
and
\bad
\frac{4 \t_1  \t_2} {\Delta s^2}  = \frac{16 \pi^2}{H^2} \left[ \frac{1}{16 \pi^2 \epsilon^2 \Sigma} - \frac{{\cal R} }{576} + \frac{{\cal R}^2 \Sigma \epsilon^2}{34560}  + {\cal O}(\epsilon^4) \right]\, .
\ead
\end{subequations}
Consequently, both \eq{G2BD00} and \eq{G2BDii} can be written as a series of positive powers of $\epsilon$.   
Then, by using the expansion identity for the hypergeometric functions \cite{Abramowitz}
\bea \label{ExpansionF}
 F \left( a,b;a+b-m;z\right) &=& \frac{\Gamma(m) \Gamma(a +b -m)}{\Gamma(a) \Gamma(b)} (1-z)^{-m} \sum^{m-1}_{n=0} \frac{(a-m)_n (b-m)_n}{n! (1-m)_n} (1-z)^n  \nonumber\\
&-& \frac{(-1)^m \Gamma(a+ b -m)}{\Gamma(a-m) \Gamma(b-m)} \sum^\infty_{n=0} \frac{(a)_n (b)_n}{n! (n+m)!} (1-z)^n \cdot \nonumber\\
&& \Bigl\{ \ln(1-z) - \psi(n+1) - \psi(n + m +1) + \psi(a+n) + \psi(b +n ) \Bigr\}\nonumber\\
\eea
it is easy to see from \eq{expz} that the first sum will mostly contain terms with negative power of $\epsilon$ and that except from the $n=0$ term in the second sum, 
all other terms will be of order $\epsilon^2$ and higher. The latter while not divergent will vanish anyways once we will take the limit 
$x_{1,2} \rightarrow x$ which corresponds to $\epsilon \rightarrow 0$ thus play no role in the final result of \eq{VEVTBD}. 
Notice that the only terms where a non-trivial dependence from $\n$ can be found are $\psi(a)$ and $\psi(b)$.

Furthermore, BD perform an expansion of the polygamma functions $\psi(a),\psi(b)$ in the dimensionless parameter
\be
\eta \equiv \frac{\cal R}{m^2} 
\ee
to order ${\eta}^2$ for small $\eta$, that gives
\bea
\psi \left( \frac{c}{2} + \nu \right) &+& \psi \left( \frac{c}{2} - \nu \right)  =  \nonumber\\
&-& \ln\left(\frac{\eta}{12} \right) + \eta \left(\xi - \frac{25}{144}+\frac{c(c-2)}{48} \right)  \nonumber\\
&+& \eta^2 \frac{-34560 \xi^2 + (1440 c^2 - 2880 c - 12000) \xi -15c^4 + 60 c^3 +210c^2 - 540c - 1027 }{69120} \nonumber\\
\eea
where $c=3,5,7$. The quantity that contains the expansion parameter is $\n = i \sqrt{\frac{12}{\eta} + 12\xi - \frac{d^2}{4} }$.
For example for $d=4$ and $c=3$, it gives:
\bad \label{psi(a) + psi(b) c=3/2}
\psi \left( \frac{3}{2} + \nu \right) + \psi \left( \frac{3}{2} - \nu \right)  = - \ln\left(\frac{\eta}{12} \right)  + \eta \left(\xi - \frac{1}{9}\right)
- \eta^2 \left( \frac{\xi^2}{2} + \frac{\xi}{9} + \frac{11}{2160} \right).
\ead
In this expansion we have then
\bad \label{Fren3/2}
\Gamma \left( \frac{3}{2} + \nu \right) \Gamma \left( \frac{3}{2} - \nu \right) & F^{\rm ren} \left( \frac{3}{2} + \nu, \frac{3}{2} - \nu;2; \frac{\chi+1}{2} \right) = \\
& \frac{m^2}{H^2} \biggl\{ \left[1 + {\eta} \left( \xi - \frac{1}{6} \right) \right]  \left[ \psi \left( \frac{3}{2} + \nu \right) + 
\psi \left( \frac{3}{2} - \nu \right) + \ln\left(\frac{\eta}{12}\right) \right] \\
& \qquad - \left( \xi - \frac{1}{6} \right) {\eta} - \frac{{\eta}}{18} - \frac{1}{2} \left( \xi - \frac{1}{6} \right)^2 \eta^2 + \frac{1}{2160}\eta^2 \biggr\}\, ,
\ead
where $F^{\rm ren}$ is equal to $F$ with the terms of $O(\epsilon^{-r})$ with $r>0$ discarded (more technically absorbed by counter-terms)
and with the terms of $O(\epsilon^r)$ having set to zero in the limit $\epsilon\to 0$. This leaves only the terms of $O(\epsilon^0)$ that appear on the right hand side of \eq{Fren3/2}.
As far as the renormalized EMT is concerned, a similar process can be followed for the other hypergeometric functions in 
\eq{GBD00}, \eq{GBDii} (for $c=5,7$) and show that after the renormalization scheme, they contain only terms of $O(\epsilon)$. 
Subsequently, in the limit $\epsilon \rightarrow 0$ both \eq{G2BD00}, \eq{G2BDii} satisfy
\bad
\lim_{\epsilon \rightarrow 0} G^{(2)}_{\rm BD;00} = 0, \qquad
\lim_{\epsilon \rightarrow 0} G^{(2)}_{{\rm BD};ii} = 0
\ead
since they contain prefactors that contribute with positive powers of $\epsilon$. This leads to \eq{Double Derivative G}. 
We now have all the pieces to compute \eq{VEVTBD}. Setting $\m = \n = 0$ for example and using \eq{Double Derivative G}, we obtain
\bea \label{VEVTBD00}
\braket{T^{\rm BD}_{00}}_{\rm BD} &=&  \lim_{\substack{x_1 \rightarrow x \\ x_2  \rightarrow x}} 
-2 \xi \biggl\{ \nabla_{i,1} \nabla_{i,2} G_{\rm BD} +  \nabla_{i,2} \nabla_{i,1} G_{\rm BD}  + \nabla_{i,1} \nabla_{i,1} G_{\rm BD} + \nabla_{i,2} \nabla_{i,2} G_{\rm BD}\biggr\} \nonumber\\
&-&  \lim_{\substack{x_1 \rightarrow x \\ x_2  \rightarrow x}}  \frac{1}{4} m^2 g_{00} G_{\rm BD}
\eea
where the remaining minimally coupled covariant derivatives along with the $G_{\mu \nu}$ via the equation of motion cancel exactly half the mass term. 
In addition, via \eq{Double Derivative G_2}, the spatial double derivatives of $G_{\rm BD}$ in the curly brackets cancel out with each other. 
The same can be shown for the $ii$ components, therefore we conclude that
\bad \label{BDTdSiso}
\braket{T^{\rm BD}_{\m \n}}_{\rm BD} &=  -\lim_{\substack{x_1 \rightarrow x \\ x_2  \rightarrow x}} \frac{1}{4} m^2 g_{\m \n} G_{\rm BD}\, ,
\ead
where now both sides are renormalized.
This, together with \eq{GBD_2} and \eq{Fren3/2}, gives our final result:
\bea\label{BDTMN}
\braket{T^{\rm BD}_{\m \n}}_{\rm BD} = &-&\frac{g_{\m \n}}{64 \pi^2} m^4 \biggl\{ \left[ 1 + \eta 
\left( \xi - \frac{1}{6} \right) \right]  \left[ \psi \left( \frac{3}{2} + \nu \right) + \psi \left( \frac{3}{2} - \nu \right) + \ln\left(\frac{\eta}{12}\right) \right] \nonumber\\
&-&  \left( \xi - \frac{1}{9}  \right) \eta  -  \eta^2 \left[\frac{1}{2} \left( \xi - \frac{1}{6} \right)^2  - \frac{1}{2160 } \right]\biggr\} \, .
\eea
This is just \eq{TmnBD} and agrees with the result given in \cite{bunch1978}. 
The above expression is an expansion in $\eta$ around the flat limit, which is the domain where the de Witt - Schwinger expansion takes place.
Regarding the quantity in \eq{BDTMN}, the two approaches agree in fact order by order in $\eta$.
In addition, the renormalized EMT satisfies the conservation equation
\bad
\nabla^\mu \braket{T^{\rm BD}_{\m \n}}_{\rm BD} = 0 \, ,
\ead
as it should. To see this notice that in the $\epsilon$-expanded expression, $\nabla^\mu \braket{T^{\rm BD}_{\m \n}}_{\rm BD} \sim \partial_\n G_{\rm BD}$ since the covariant derivative acts on 
scalar quantities. In addition the space-time dependence in $G_{\rm BD}$ is exclusively via $\epsilon$, coming to leading order form terms of order $\epsilon^2$, whose derivative 
vanishes in the $\epsilon\to 0$ limit.

Another issue is the fact that the function $ \psi \left( \frac{3}{2} - \nu \right)$ is singular for the Bunch-Davies mode function with $\n=\frac{3}{2}$.
This is however consistent since away from the conformal limit the classical backreaction protects $\n$ from reaching its conformal value 3/2.
The non-trivial conformal point\footnote{This is a point where $m^2+\xi{\cal R}=0$ so that $\m^2=0$ in \eq{ActionS}. } that is the $3d$ flat horizon, is however singular. 
This is similar to the RG flow of scalar theories in $d=4$ that develop a 
non-trivial Wilson-Fisher fixed point in dimensional regularization, where the system becomes three-dimensional. In this IR limit, QFT correlators develop 
a logarithmic singularity, as they transit abruptly from obeying Callan-Symanzik equations, to obeying Dilatation Ward Identities.
This analogy suggests perhaps that the horizon should be better described by a CFT if not by a carefully regularized limit.

\section{Classical and Bunch-Davies backreactions} \label{Zero-temp Backreaction}
Suppose that in an empty 4d dS spacetime, we spontaneously turn on a test field $\phi$ which can be written as
\bad
\phi(\t,\tx) =c_1 \phi_{\rm cl}(\t) +  c_2 \phi_{\rm BD} (\t,\tx)
\ead 
with $\phi_{\rm cl}(\t)$ and $\phi_{\rm BD}(\t,\tx)$ being the solutions to the spatial-independent classical and Bunch Davies eoms correspondingly, and $c_1,c_2$ parameters that satisfy $c_1 = c_2 = 1$ in natural units. This sudden appearance of the field will trigger, through the Einstein eq. \eq{EinEq.}, a backreaction which should deform both the spacetime geometry and the field as a result. Fortunately enough, if one proceeds to compute the EMT as it is given from \eq{Tmn}, they will eventually encounter the mixed expectation values $\braket{\phi_{\rm cl} \phi_{\rm BD}}$ which we set equal to zero, due to them containing a single ladder operator coming from the expansion \eq{scalarfieldi}. Thus, the RHS of \eq{EinEq.} can be broken down into a classical and Bunch Davies part:
\bad \label{genTmn}
\braket{T_{\mu \nu}}_{\rm BD} = c_1^2 T^{\rm cl}_{\mu \nu} + c_2^2 \braket{T^{\rm BD}_{\mu \nu}}_{\rm BD} 
\ead
and since both the classical component and the dS VEV of the EMT are spatial independent, the general EMT should be as well. This enables us to assume as in \sect{thermal backreaction} that the new deformed metric will be of the FRW form:
\begin{subequations}
\bad
ds^2 = \tilde a^2(\t) \left(- d \t^2 + d \tx^2 \right)\, ,
\ead
where the new scale factor can be written as
\bad
\tilde a(\tau) = a(\tau) + \delta a(\tau) 
\ead
\end{subequations} 
with our goal being to compute the correction $\delta a$ at least to first order in the Newton constant $G$. having defined
\bad
\tilde H(\tau) = \frac{\tilde a'(\tau)}{\tilde a^2(\tau)}
\ead
as the analogue of the Hubble constant for the deformed metric, allows us to reduce the Einstein equation to:
\begin{subequations} \label{ODEsystem}
\begin{gather} \label{Fried1}
\tilde H^2 = H^2  + \frac{8 \pi G}{3} \frac{1}{\tilde a^2 } \braket{T_{00}}_{\rm BD} \\
\frac{(\tilde a')^2}{\tilde a^4} - 2 \frac{\tilde a''}{\tilde a^3}  =- 3 H^2 +8 \pi G \frac{1}{\tilde a^2} \braket{T_{11}}_{\rm BD} \label{Fried2}\, ,
\end{gather}
the corresponding Friedmann equations. Here $T_{00}$ and $T_{ii}$ are the time and spatial components of the EMT which can be calculated once one solves
\bad \label{ClassEom}
\phi'' + 2 \frac{\tilde a'}{\tilde a} \phi' + \tilde a^2 \mu^2 \phi = 0 \, ,
\ead
i.e. the classical eom. Here we have suppressed the spatial-derivative terms and
\bad \label{BDeom}
u''_\tmk + 2\frac{\tilde a'}{\tilde a} u'_\tmk + \left( \tk^2 + \tilde a^2 \mu^2 \right)u_\tmk = 0
\ead
\end{subequations}
is the analogue of \eq{Covarianteom} for the deformed metric.

The general solution to \eq{ClassEom} for a dS background is
\bad
\phi = \phi_0 \left(\tau^{\frac{3}{2} + \nu} + {\cal C} \tau^{\frac{3}{2} - \nu} \right)
\ead
with $\phi_0$ the vacuum state of the field. In order for the field to retain its proper time-dependence when one sets $\nu  = 3/2$ 
and to have no translation of the vacuum energy density by a finite quantity, we require ${\cal C} = 0$ so that
\bad
\phi = \phi_0 \tau^{\frac{3}{2} + \nu}  \ . 
\ead
Note that in natural units, the classical dimension of $\phi_0$  is $[\phi_0] = - \nu -1/2$ so that the field $\phi$ has the correct dimension. Then, the $00$ and $ii$ components of the EMT are
\begin{subequations} \label{ClassTmncomp}
\begin{gather}
T_{00} = \frac{\phi^2_0}{2} \left[\frac{12}{\eta} + \left(\frac{3}{2} + \nu  \right)^2- 12 (1 + \nu) \xi \right] \tau^{1+ 2\nu} \equiv 6 \phi^2_0 \tau^{1+ 2\nu} \frac{A(\eta)}{\eta} 
\end{gather}
and
\begin{gather}
T_{ii} = \frac{\phi^2_0}{2} \left[-\frac{12}{\eta} - 6 \xi +  \left( \frac{3}{2} + \nu \right)^2 (1 -4 \xi) + 4 \xi \left(\frac{1}{2} - \nu \right) \left( \frac{3}{2} + \nu \right) \right] \tau^{1+ 2\nu} \equiv 6 \phi^2_0  \tau^{1+ 2\nu} \frac{B(\eta)}{\eta}
\end{gather}
\end{subequations} 
respectively with
\begin{subequations}
\bad \label{A(eta)}
A(\eta) = 1 + \eta \left[ \frac{1}{12} \left(\frac{3}{2} + \nu  \right)^2-  (1 + \nu) \xi \right]
\ead
and
\bad
B(\eta) = -1 + \eta \left[  \frac{\xi}{3}  \left(\frac{1}{2} - \nu \right) \left( \frac{3}{2} + \nu \right) - \frac{\xi}{2}  + \frac{1}{12} \left( \frac{3}{2} + \nu \right)^2 (1 -4 \xi)  \right]  \, . 
\ead
\end{subequations}
Note here that the above formulae satisfy the identity $A+B = \frac{1}{3} (3 + 2 \n) A $. Substituting the above into either Friedmann eq. (\eq{Fried1} or \eq{Fried2}) for $c_1=1, c_2=0$, the resulting ODE up to first order in $G$, are solved by:
\bad
\tilde a_{\rm cl}(\tau) = a(\tau) \left(1 - \frac{4 \pi}{ (2 + \nu)} G \phi^2_0  \tau^{3+ 2\nu} \frac{A(\eta)}{\eta} \right) 
\ead
which in return gives
\bad
\tilde H_{\rm cl} = H \left(1 + 8 \pi  G \phi^2_0  \tau^{3 + 2 \nu}  \frac{A(\eta)}{\eta}\right) \ .
\ead
Such corrections were also found in \cite{Starobinsky1980}. 

Assuming a fixed dS background allows us to directly use \eq{BDTMN} which we can also substitute 
into either \eq{Fried1} or \eq{Fried2} with $c_1=0$ and $c_2 =1$. Then, the solution to first order is
\bad
\tilde a_{\rm BD}(\t)  = a(\tau) \left( 1 - \frac{1}{48\pi} G H^2 \frac{ C(\eta)}{\eta^2} \right) 
\ead
with
\bad
C(\eta) &= \biggl\{ \left[ 1 + \eta 
\left( \xi - \frac{1}{6} \right) \right]  \left[ \psi \left( \frac{3}{2} + \nu \right) + \psi \left( \frac{3}{2} - \nu \right) + \ln\left(\frac{\eta}{12}\right) \right] \\
& \qquad -  \left( \xi - \frac{1}{9}  \right) \eta  -  \eta^2 \left[\frac{1}{2} \left( \xi - \frac{1}{6} \right)^2  - \frac{1}{2160 } \right]\biggr\}
\ead 
a real function of $\eta$. 

Finally for the combined case ($c_1=c_2 = 1$), due to the non-existence of any mixing between the classical and BD contributions, the general deformed scale factor is seen to be
\bad
\tilde a(\t) =  a(\tau) \left(1 -  \frac{4 \pi}{ (2 + \nu)} G \phi^2_0  \tau^{3+ 2\nu} \frac{A(\eta)}{\eta} - \frac{1}{48 \pi} G m^2 \frac{C(\eta)}{\eta}  \right) \, ,
\ead
which can alternatively be read as a time-dependent correction to the dS Hubble constant
\bad \label{BDbackreact}
\tilde H = H \left(1 + 8 \pi G \phi^2_0  \tau^{3 + 2 \nu} \frac{A(\eta)}{\eta} +  \frac{1}{48 \pi} G m^2 \frac{C(\eta)}{\eta}  \right) \ .
\ead
Then, we can rewrite the above result as 
\bad 
\tilde H = H \left(1 +  8 \pi G \phi^2_0  \tau^{3 + 2 \nu} \frac{A(\eta)}{\eta} +  \frac{\hbar}{ 48 \pi} \frac{m^2}{M^2_{\rm pl}} \frac{C(\eta)}{\eta} \right) \ ,
\ead
where we expressed Newton's constant in terms of the Planck mass $M_{\rm pl}$, $G= \hbar/ M^2_{\rm pl}$, so that the quantum correction obtains a factor of $\hbar$.

\printbibliography 

\end{document}